\documentclass[twocolumn,showpacs,preprintnumbers,amsmath,amssymb]{revtex4}

\usepackage{graphics}
\usepackage{graphicx}
\usepackage{amsmath}
\usepackage{amsfonts}
\usepackage{amssymb}
\usepackage{pstricks}

\newcommand{\dev}{\tilde}
\newcommand{\mat}{\text}
\newcommand{\ex}{\hat{\vec{e}}_x}

\newcommand{\ea}[1]{e^{\mat{L}#1}}

\newcommand{\ez}{\hat{\vec{e}}_{v}}

\newcommand{\er}{\hat{\vec{e}}_{r}}

\renewcommand{\vec}[1]{\mathbf{#1}}

\newcommand{\Wi}{\text{Wi}}

\begin{document}

\title{Dynamics of a trapped Brownian particle in shear flows}

\author{Lukas Holzer$^{1}$, Jochen Bammert$^{1}$, Roland Rzehak$^{2}$, Walter Zimmermann$^{1}$}

\affiliation{
$^{1}$Theoretische Physik I, Universit\"at Bayreuth, D-95440 Bayreuth \\ 
$^{2}$Forschungszentrum Dresden, D-01328 Dresden
}

\date{Received: \today / Revised version: \today}

\begin{abstract}
The Brownian motion of a particle in a harmonic potential, which is 
simultaneously exposed either to a linear shear flow or to a plane Poiseuille
flow is investigated. In the shear plane of both flows the probability distribution of
the  particle becomes anisotropic
 and the dynamics is changed in a characteristic manner compared
to a trapped particle in a quiescent fluid. 
The particle distribution takes either an elliptical or a
parachute shape or a superposition of both depending  
on the mean particle position in the shear plane. 
Simultaneously, shear-induced cross-correlations 
between particle fluctuations along orthogonal directions
in the shear plane are found.
They are asymmetric in time.
In Poiseuille flow thermal
particle fluctuations perpendicular 
to the flow  direction 
in the shear plane induce a shift of the particle's mean position 
away from the  potential minimum.
Two complementary methods are suggested
to measure shear-induced cross\--correlations between
particle fluctuations along
orthogonal directions.
\end{abstract} 


\maketitle

\section{Introduction}
\label{intro}

The Brownian motion of particles in a fluid is of central importance 
in chemical and biological physics as well as in material science and engineering
\cite{Einstein:1905.1,Dhont:96,Stone:2001.1,Ottino:2004.1}. 
Despite the long history of Brownian motion, especially in quiescent fluids,
 our understanding of thermally induced particle dynamics  in flows 
is still far from complete.

Moreover, neutral colloidal particles moving relatively to each other interact via the fluid 
and these hydrodynamic interactions can cause a complex collective behavior \cite{Dhont:96,Holzer:2006.1,Bechinger:2006.1}.
In shear flows little is known 
about the dynamics of
Brownian particles and the hydrodynamic interaction effects
in spite of their fundamental relevance and importance in microfluidic applications.
The 
Taylor dispersion  \cite{TaylorGI:1953.1} and fluid mixing issues \cite{Ottino:2004.1,Steinberg:2001.2} 
are well known examples where fluctuations of particles and their hydrodynamic interaction
effects in simple shear and Poiseuille flow play an important role.
The interplay of shear gradients and thermal motion
of polymers leads, even at low values of the Reynolds number, to the rich
dynamics of polymers \cite{Chu:1997.1}, the so-called molecular
individualism \cite{deGennes:1997.1}. Polymers tumbling
in a shear flow cause elastic turbulence even 
in diluted polymer solutions \cite{Steinberg:2000.1} and spectacular mixing properties \cite{Steinberg:2001.2} in microchannels. 

In shear flows it 
is the contribution $(\vec{u} \cdot \nabla) \vec{u}$ of the flow field $\vec{u}(\vec{r})$
to the Navier-Stokes  equation which
causes  interesting transient phenomena
near the onset
of turbulence \cite{SGrossmann:2000.1}, as well as 
amplifications of velocity fluctuations and their cross-correlations along
and perpendicular to straight streamlines
 \cite{Eckhardt:2003.1,Oberlack:2006.1}. 
A cross-correlation is also expected 
between orthogonal particle-fluctuations
in the shear plane, because random jumps of a particle
between neighboring  streamlines of different velocity
lead to a change in the particle's velocity 
and displacement  along the streamlines, similar as via fluctuations.
Nevertheless, there was no direct observation
of these cross-correlations until recently \cite{Ziehl:2009.1}.
Here we present the theoretical background for their determination.

The stochastic dynamics of free single spherical particles in linear shear flows
have been studied in terms of the hydrodynamic fluctuation theory
 \cite{Bedeaux:1995.1}, and in combined Langevin and Fokker-Planck
approaches, by taking inertia into account \cite{Brady:04,Drossinos:05,Sancho:1979.1}. Even effects of non-equilibrium thermodynamics
were included in Ref.~\cite{Holek:01}.
Experiments for detecting shear-induced cross-correlations 
between perpendicular random displacements of free particles 
were described in Ref.~\cite{vdVen:1980.1,Leal:1980.1}. Shear induced
cross-correlations between perpendicular fluid-velocity fluctuations
and perpendicular fluctuations of particles
are expected to be strongly asymmetric in time
  \cite{Eckhardt:2003.1,Drossinos:05,Holzer:2009.2}.
In dynamic light-scattering experiments certain 
aspects of this issue 
were observed
indirectly \cite{vdWater:1998.1}, however,
a direct measurement and characterization of related 
particle fluctuations in shear flows remained an open question.

Direct observations of particle fluctuations 
at the meso\-scale became possible only quite recently
by using optical tweezers. This rather young
technique is
a powerful experimental method for investigating the
motion of a small number of particles \cite{Grier:2003.1,Ashkin:1986.1,Chu:1991.1}, which
contributes substantially to our understanding of the dynamics of
particles and to a number
of innovative applications. 
These include the inspiring  studies on single polymers \cite{Chu:1994.2,Chu:1995.1,Brochard:1995.1,Larson:1997.1,Rzehak:99.2,Rzehak:00.1,Kienle:01.1,Rzehak:02.1}, the detection of
anti-correlations between hydrodynamically interacting Brownian particles
by Femto-Newton measurements \cite{Quake:99.1}, the propagation of hydrodynamic interactions  \cite{Bartlett:2002.1}, wall effects on Brownian motion \cite{Grier:2001.1,Grier:2000.1}, 
short-time inertial response of viscoelastic fluids \cite{Schmidt:2005.1},
two-point microrheology \cite{Weitz:2000.1}, 
anomalous vibrational dispersion \cite{Quake:2006.1}, particle sorting
techniques \cite{Grier:2002.1,Bammert:2008.1,Bammert:2009.1,Dholakia:2003.1}, and a number of further investigations in 
microfluidics. The laser tweezer technique has also been applied to determine the force elongation
relation of biopolymers \cite{Simons:BPJ70-96-1813} or the effective pair
potential in colloidal suspensions \cite{Crocker:JCIS179-96-298}.

Stochastic motions of a free particle 
moving along the streamlines of a sheared fluid and of a 
particle trapped in the minimum of a potential,
while exposed to a shear flow, have common characteristic signatures.
Since the trapped particles 
are more suited for a thorough statistical analysis of
its Brownian dynamics,  we present calculations
for particles trapped by a harmonic potential and exposed to
either a linear shear flow or to a plane Poiseuille flow.
Our analytical results show that shear flow causes characteristic signatures in the time dependence of the 
cross-correlation between particle displacements along orthogonal directions in the shear plane as well as an inclined elliptical particle distribution.
Part of these results have already been applied
and confirmed in a recent experimental study
\cite{Ziehl:2009.1}.

For our calculations we
utilize a Langevin model for the particle motion, where  stochastic forces with different
statistical properties may be used.
Stochastic forces acting on suspended Brownian particles   
are caused by velocity fluctuations of the surrounding fluid. 
In a quiescent fluid the fluctuations of 
orthogonal velocity components are uncorrelated
in the bulk \cite{LanLifVI}.  Assuming 
such uncorrelated fluid-velocity fluctuations, and therefore
uncorrelated stochastic forces in the 
 particle Langevin model, we show how  shear flow induces
cross-correlations between particle fluctuations along
orthogonal directions. 
Conversely, we show how the amplitudes of the stochastic forces acting on the particle can be determined by measuring the static correlations of the particle fluctuations.

In section~\ref{S:eq_of_motion} 
the model equations of the 
Brownian particle motion and their formal analytical solutions are presented. 
The static correlation functions for the particle's position and velocity
fluctuations are derived in section~\ref{S:fluct}, where also
the corresponding distributions are calculated in terms of the static correlations and under the assumption of Gaussian particle fluctuations.
In addition, the ratio between the principal axes of both distributions are determined as well as the
angle enclosed   by each major axis and the flow direction.
In Poiseuille flow the second derivative of the flow profile,
as well as the fluctuations perpendicular to the flow lines,
cause a shift of the particle's mean position in the potential via  $\Delta {\bf u}({\bf r}) \not =0$ .
The latter contribution is usually not taken into account, if the effective
particle radius is determined
via Fax\'en's law from the particle displacement.
In section~\ref{S:results} we present and discuss these results for
the special cases of linear shear flow and plane Poiseuille flow.
In addition  we compare the analytical results with numerical simulations of the
Langevin equation given in section~\ref{S:eq_of_motion} and we suggest experiments to measure some of the flow-induced effects determined in 
this work. 
The article closes with a discussion and possible further applications in section~\ref{conclusion}.

\section{Equations of motion and their solutions}
\label{S:eq_of_motion}

We consider a Brownian particle of mass $m$ and effective radius $R$ 
suspended at the position $\vec{r}=(x,y,z)$ in a flow field 
with parallel streamlines in the $x$-direction, $\vec{u}(\vec{r}) = u_x(y) \ex$.
We assume a velocity field 
\begin{align}
	u_x(y)=(a + by + cy^2) \,,
	\label{E:flow}
\end{align}
which corresponds for $b=c=0$ to a uniform flow,
for $a=c=0$ to a linear shear flow with shear rate $b$, and 
for $c=-\frac{a}{l^2}$, $b=0$ to a plane Poiseuille flow between two parallel walls 
at a distance $2l$.
The  particle is trapped by a harmonic potential with its minimum at $\vec{r}_0=(x_0,y_0,z_0)=(0,0,0)$,
\begin{align}
	U(\vec{r})= \frac{k}{2}\vec{r}^2 \label{E:potential}\,.
\end{align}
The resulting linear restoring force is given by
\begin{align}
\label{E:springforce}
	\vec{F}^p=-\nabla U= -k \vec{r} \,,
\end{align}
in terms of the force constant $k$.
Such a potential acting on a colloidal particle may be
realized by an optical tweezer \cite{Grier:2003.1}.

A particle moving with the speed  $\vec{v}=\dot {\bf r}$ in a flow of
velocity ${\bf u}$ experiences, according to Stokes' law, a hydrodynamic drag
force ${\bf F}^h=6 \pi \eta R (\vec{u} - \vec{v})$ 
proportional to the effective radius $R$, to the shear
viscosity $\eta$ and 
to the difference $\vec{u} - \vec{v}$ between the
velocity of the particle and the local flow velocity \cite{Stokes:1850.1,LanLifVI}. 
If the flow velocity $\vec{u}(\vec{r})$ is a nonlinear function of the spatial coordinates,
 as in the case of a plane Poiseuille flow in Eq.~\eqref{E:flow}, one has according to
Fax\'en's theorems \cite{Faxen:27.1} an additional contribution to the drag force.
This contribution includes the Laplacian of the velocity field and has in terms
of the Stokes friction coefficient, $\zeta=6\pi\eta R$, the  form:
\begin{align}
	\vec{F}^h &= - \zeta \left( \dot{\vec{r}} - \vec{u}( \vec{r} )
			- \frac{R^2}{6} \Delta \vec{u}( \vec{r}) \right) 
\,. \label{E:Faxen}
\end{align}
The Laplacian of the flow field in Eq.~\eqref{E:flow},
with the  abbreviation $\bar{a}=a+\frac{R^2}{3}c$, gives the following 
expression for the hydrodynamic drag force,
\begin{align}
	\vec{F}^h &= - \zeta \dot{\vec{r}} + \zeta(\bar{a} + b y + c y^2)\ex \,.
	\label{E:force_hydro}
\end{align}

The stochastic motion of a Brownian particle is caused by 
the  fluctuations 
 $\tilde {\bf u}$
of the fluid velocity-field ${\bf u}({\bf r},t)$ \cite{Reichl:1998}. The effects of
$\tilde {\bf u}$ on a particle 
can be taken into account in a Langevin model 
by a random force $\vec{F}^b(t)$.  
In uniform flows, namely with $b = c = 0$ in
Eq.~\eqref{E:flow}, the cross-correlations of the velocity
fluctuations of the fluid, $\langle \dev{u} _i \dev{u}_j \rangle$,
lead to a vanishing cross-correlation of the
random forces, $\langle F_i^b F_j^b \rangle=0 \,, (i \neq j)$.
The shear-induced 
contributions to $\tilde u_i$ and $F_i^b$ are a matter of
current research \cite{Eckhardt:2003.1,Oberlack:2006.1}.

In our model we assume a Gaussian distribution of $\vec{F}^b(t)$ 
with vanishing correlation time and mean value
\begin{align}
	\langle F_i^b(t) \rangle &= 0 \,, \notag \\
	\langle F_i^b(t) F_j^b(t') \rangle &= f_{ij} \delta(t-t') &
	\text{and $i,j \in {x,y,z}$}\,.
	\label{E:fluctcorr}
\end{align}
For the moment we leave the fluctuation matrix $f_{ij}$ unspecified, except to note that,
according to time-translation- and time-reversal-invariance, 
it is symmetric. 
In uniform flows, namely 
with $b=c=0$ in Eq.~\eqref{E:flow}, the matrix $f_{ij}$ is diagonal with $f_{ii}=2 k_BT \zeta$ \cite{LanLifIX} as mentioned before, whereas
in a shear flow the magnitude of the non-diagonal elements of $f_{ij}$ 
depends on the shear rate, but shear-induced contributions are expected to be small
\cite{Eckhardt:2003.1,Holzer:2009.2,Bedeaux:1995.1}.
However, we allow non-diagonal elements of $f_{ij}$ for the moment 
to show in  section~\ref{S:results} how these non-diagonal elements 
may be determined by measurements of the velocity fluctuations of the
particle.

The net force acting on the particle,
\begin{align}
	\vec{F} = \vec{F}^h + \vec{F}^p + \vec{F}^b(t) \,,
\end{align}
together with Newton's law gives the Langevin equations of motion
for the translational degrees of freedom of the particle
\begin{subequations}
\begin{align}
m\ddot{x} &= - \zeta \dot{x} - k x + \zeta ( \bar{a} + by +c y^2 ) +
		F^b_x(t)\,, \\
m\ddot{y} &= - \zeta \dot{y} - k y + F^b_y(t) \,, \\
m\ddot{z} &= - \zeta \dot{z} - k z + F^b_z(t) \,. 
\end{align}
\label{E:lang_equation}
\end{subequations}
Introducing the vectors $\vec{X}=(x,v_x)$, $\vec{Y}=(y,v_y)$ and $\vec{Z}=(z,v_z)$ one may express the second order differential equations
\eqref{E:lang_equation} in terms of a system of coupled first order equations
\begin{subequations}
\begin{align}
	\dot{\vec{X}} &= \mat{L} \vec{X} + \frac{F^b_x(t)}{m} \ez 
					 + 2 \beta ( \bar{a} + by +c y^2 ) \ez
\,, 
\\
	\dot{\vec{Y}} &= \mat{L} \vec{Y} + \frac{F^b_y(t)}{m} \ez \,, 
\\
	\dot{\vec{Z}} &= \mat{L} \vec{Z} + \frac{F^b_z(t)}{m} \ez \,.
\end{align}
\label{E:lang_vec}
\end{subequations}
Herein we have introduced the matrix
\begin{align}
        \mat{L} &=  \begin{pmatrix}
                         0   	& ~1 \\
                        -\omega^2~& -2\beta~
                    \end{pmatrix}, 
\end{align}
the damping constant $\beta=\frac{\zeta}{2m}$, the squared frequency $\omega^2=\frac{k}{m}$
and the velocity unit-vector $ \ez=  ( 0\,,\, 1)$.
The rotational motion of the particle, which may provide further
corrections to the leading order fluctuation effects discussed here, is not taken into
account in this work.

The general solutions of the equations of motion~\eqref{E:lang_vec}
in terms of the initial conditions $\vec{X}(0)$, $\vec{Y}(0)$ and $\vec{Z}(0)$
are given by 
\begin{subequations}
\label{E:sol_r}
\begin{align}
        \vec{X}(t)&= \ea{t}\vec{X}(0) 
			+ \int_0^t d\tau \, \ea{(t-\tau)}\,
					\frac{F^b_x(\tau)}{m}~ \ez 
	\label{E:sol_X} \\
		  &\quad + 2 \beta \int_0^t d\tau \, \ea{(t-\tau)} 
					( \bar{a} + by +c y^2 ) ~\ez \,,
	\notag \\
        \vec{Y}(t)&= \ea{t}\vec{Y}(0) + \int_0^t d\tau \, \ea{(t-\tau)}\,
			\frac{F^b_y(\tau)}{m}~\ez  \,, 
	\label{E:sol_Y} \\
	\vec{Z}(t)&= \ea{t}\vec{Z}(0) + \int_0^t d\tau \, \ea{(t-\tau)}\,
			\frac{F^b_z(\tau)}{m}~\ez  \,.
\label{E:sol_Z}
\end{align}
\end{subequations}
This system of coupled equations is the starting point of the following analysis,
where the statistical properties of the particle's motion 
are characterized by the correlations of its position
and velocity.

\section{Distribution functions and static correlations}
\label{S:fluct}

By taking the averages of Eqs.~\eqref{E:sol_r} and using the vanishing mean of the stochastic forces in Eq.~\eqref{E:fluctcorr} we see that
the mean velocity 
of a particle in a harmonic potential vanishes, $\langle {\vec{v}} \rangle=0$.
In the case $\vec{r}_0=0$ the mean value of the deviations of the particle's position
from the potential minimum in the directions perpendicular to the flow vanish too, $\langle y \rangle = \langle z \rangle =0$ .
However, the mean particle displacement in the direction of the flow given by Eq.~\eqref{E:flow} is non-zero:
\begin{align}
	\langle x \rangle = \frac{\zeta}{k} \bar{a}
			      +  \frac{f_{yy}}{2 k^2} c
			  			\,. \label{E:mean_x}
\end{align}
This equation is discussed in more detail in section~\ref{S:results} for specific flows.
Since the spring constant $k$ of the potential enters
in different powers in equation~\eqref{E:mean_x}, this formula might be
employed for the experimental determination 
of the effective fluctuation magnitude $f_{yy}$.

With a combination of the coordinates of the particle and its
velocities to a single $6$-dimensional vector, $\vec{q}=(\vec{r},\vec{v})=(x, y ,z, v_x, v_y,
v_z)$, the probability distribution function of the particle, $\cal P({\bf q})$,
may be formulated  in a compact form in terms of
the deviations $\dev{\vec{q}}=\vec{q}-
\langle \vec{q} \rangle = (\dev{x},\dev{y},\dev{z},\dev{v}_x, \dev{v}_y,
\dev{v}_z)$  from the mean value  $\langle {\bf q} \rangle $. 
If the fluctuations $\dev{\vec{q}}$ are linear functions of Gaussian distributed stochastic forces, which is the case for uniform and linear shear flows, 
they can be expected to be themselves Gaussian variables \cite{Kampen}
and can be described by  a Gaussian distribution as follows:
\begin{align}
\mathcal{P}(\dev{\vec{q}}) &= \frac{1}{\sqrt{(2\pi)^6 \det(\mat{C})}}
\exp\left(
	-\frac{1}{2}\dev{\vec{q}}^T \mat{C}^{-1}\dev{\vec{q}}
    \right)
\, . \label{E:gauss_dist}
\end{align}
Here the covariance matrix 
$\mat{C} = \langle \dev{\vec{q}} \dev{\vec{q}}^T \rangle$
is used, which includes second order moments for
the coordinates and the velocities at equal times (static correlations). The magnitudes of the
elements $\mat{C}_{ij}$ depend on the correlation amplitudes
 of the stochastic forces,  $f_{ij}$, and can be measured in experiments.
Consequently, one may reconstruct the
stochastic forces from the measurements, as discussed
later.

For Poiseuille flow the relations between  particle fluctuations 
and  stochastic forces are nonlinear due to the contribution $\propto c y^2$
in Eq.~(\ref{E:sol_X}). Therefore, 
the particle distribution $\mathcal{P}(\dev{\vec{q}})$ is 
not necessarily Gaussian. However, in the course of this work we use
the formula in Eq.~(\ref{E:gauss_dist}) also for particles in a potential,
which are exposed to a Poiseuille flow, but with the covariance matrix 
$\mat{C}_{ij}$ now determined in terms of the parameters for the
Poiseuille flow. 
The validity of this heuristic approach 
will be tested later in Sec.~\ref{S:poiseuille} by numerical simulations.

\subsection{Covariance matrix and angular momentum}

The covariance matrix $\mat{C}$ consists of four $3 \times 3$ submatrices
\begin{align}
\mat{C}&=
        \begin{pmatrix}
        \mat{C}_{rr} & \mat{C}_{rv}\\
        \mat{C}_{vr} & \mat{C}_{vv}
        \end{pmatrix}
\,,
\end{align}
that describe the autocorrelation for the positions $C_{rr}$ and
velocities $C_{vv}$ and their cross-correlations $C_{vr}$ and $C_{rv}$
at equal times. The
matrices  
\begin{subequations} \label{E:cova_ma}
\begin{align}
\mat{C}_{rr} &=
        \begin{pmatrix}
        \langle \dev{x}^2 \rangle & \langle \dev{x}\dev{y} \rangle & ~~~0~~~ \\
        \langle \dev{y}\dev{x} \rangle & \langle \dev{y}^2 \rangle & ~~~0~~~ \\
	~~~0~~~ & ~~~0~~~ & \langle \dev{z}^2 \rangle\\
        \end{pmatrix}\,,\label{E:crr}\\
\mat{C}_{vv} &=
        \begin{pmatrix}
        \langle \dev{v}_x^2 \rangle   &  \langle \dev{v}_x\dev{v}_y \rangle &
~~~0~~~\\
        \langle \dev{v}_y \dev{v}_x \rangle  &  \langle \dev{v}_y^2  \rangle &
~~~0~~~\\
	~~~0~~~  &  ~~~0~~~ & \langle \dev{v}_z^2 \rangle\\
        \end{pmatrix}\,, \label{E:cvv}\\
\mat{C}_{rv}= \mat{C}^T_{vr} &=
        \begin{pmatrix}
         ~~~0~~~ & \langle \dev{x} \dev{v}_y \rangle & ~~~0~~~\\
         \langle \dev{y} \dev{v}_x \rangle~~  & ~~~0~~~ & ~~~0~~~ \\
	 ~~~0~~~ & ~~~0~~~ & ~~~0~~~ \\
\end{pmatrix}\,, \label{E:crv}
\end{align}
\end{subequations}
may be calculated in terms of the expressions given by Eqs.~(\ref{E:sol_r}).

Under the Gaussian assumption, the four-point correlations of the stochastic forces, which occur due to the
quadratic contribution $y^2$ of the flow in Eq.~\eqref{E:sol_X}, can be decomposed into two-point
correlation functions according to  Wick's theorem \cite{Kampen}:
\begin{align}
        \langle F^b_i(t_1) F^b_j(t_2) F^b_k(t_3) &F^b_l(t_4) \rangle  \notag \\
                        =& ~~~~f_{ij}f_{kl} \delta(t_2-t_1)\delta(t_4-t_3)
\notag\\
                        &  + f_{ik}f_{jl} \delta(t_3-t_1)\delta(t_4-t_3)
\notag\\
                        &  + f_{il}f_{jk} \delta(t_4-t_1)\delta(t_3-t_2) \,.
			\label{E:f_moments}
\end{align}
Introducing the relaxation time of the particle in a 
harmonic potential
\begin{align}
\label{E:tau}
\tau_p=\frac{\zeta}{k}\,,
\end{align}
 the non-zero components of the covariance matrices may be written in
terms of the amplitudes of the stochastic forces and the flow parameters 
by the following set of equations:
\begin{subequations}
\label{E:shearall}
\begin{align}
\langle \tilde v_y^2 \rangle &= \frac{1}{4} \frac{f_{yy}}{m^2 \beta}~,\qquad
\langle \tilde y^2 \rangle =  \frac{\langle \tilde v_y^2 \rangle}{\omega^2}~,\\
\langle \tilde v_z^2 \rangle &= \frac{1}{4} \frac{f_{zz}}{m^2 \beta},\qquad
\langle \tilde z^2 \rangle =  \frac{\langle \tilde v_z^2\rangle }{\omega^2}~,\\
\langle \tilde v_x \tilde v_y \rangle &= \langle \tilde v_y \tilde v_x \rangle
= \frac{1}{4} \frac{f_{xy}}{m^2 \beta}~,
\label{E:vxvy}\\
\langle \dev{x}\dev{y} \rangle
        &=   \langle \dev{y}\dev{x} \rangle=
\frac{\langle \dev{v}_x \dev{v}_y \rangle}{\omega^2} 
	   + \frac{1}{2} b \tau_p \langle \dev{y}^2 \rangle 
	   \,, \label{E:shear_rxry} \\
\langle \dev{x}\dev{v}_y \rangle &= - \langle \dev{y}\dev{v}_x \rangle
        =   -  \frac{b}{2} \langle \dev{y}^2 \rangle 
	  \,, \label{E:shear_rxvy}\\
\langle \dev{v}_x^2 \rangle
        &=  \frac{1}{4}\frac{f_{xx}}{m^2\beta}
		+ \frac{1}{2} b^2 \langle \dev{y}^2 \rangle +
\frac{ 8 c^2\tau_p^2 \omega^2}{3(1+2 \tau_p^2 \omega^2)} \langle \tilde y^2\rangle^2  
	\,, \label{E:shear_vx2} \\
\langle \dev{x}^2 \rangle
        &=   \frac{\langle \dev{v}_x^2 \rangle}{\omega^2} 
	   + \tau_p b
		\langle \dev{x} \dev{y} \rangle +\frac{2}{3} \tau_p^2 c^2 
\langle \tilde y^2\rangle ^2
	   \,. \label{E:shear_rx2} 
\end{align}
\end{subequations}
The non-diagonal elements of  $C_{vv}$, namely
  the correlations of the velocity fluctuations
of the particle, 
$\langle \dev{v}_x\dev{v}_y \rangle$ and 
$\langle \dev{v}_y\dev{v}_x \rangle$, are directly proportional
to the amplitude $f_{xy}$ of the fluctuations given by Eq.~\eqref{E:fluctcorr}. 
In contrast to this result for a fluctuating particle in a potential,
one finds for a free particle in shear flow 
finite values for the cross-correlations $\langle \dev{v}_x\dev{v}_y \rangle$ and 
$\langle \dev{v}_y\dev{v}_x \rangle$ even in the case $f_{xy}=0$ \cite{Drossinos:05}. 
However, as shown appendix \ref{App_corr}, in the presence of a harmonic potential 
these correlations decay on
a time scale $1/(2\beta)$, which is very short for an overdamped particle
motion. In the case of a  weak laser tweezer potential the particle relaxation
time $\tau_p=\zeta/k$ becomes rather large and
 one obtains according to Eq.~(\ref{vxvyG}) and Eq.~(\ref{G2exp}) 
for $f_{xy}=0$ a contribution to the velocity correlation
$\langle \dev{v}_x(t)\dev{v}_y(t) \rangle \propto \exp[-2t/\tau_p]~ f_{yy}$,
which decays slowly and gives in the limit of a vanishing potential
 ($k \to 0,\tau_p \to \infty$) a constant contribution 
to the velocity-velocity correlation, which agrees with that in Refs.~\cite{Drossinos:05}
for a free particle,
 c.f. appendix \ref{App_corr}.
Therefore, finite values 
of $\langle \dev{v}_x\dev{v}_y \rangle$ and $\langle \dev{v}_y\dev{v}_x \rangle$ 
measured for  particles trapped in  a potential 
are a direct indication of 
cross-correlations of the stochastic 
forces along orthogonal directions, $\langle F^b_x F^b_y \rangle \neq 0$.

 For $b \neq 0$ the contributions to the non-diagonal 
elements
$\langle \dev{x}\dev{y} \rangle$ and 
$\langle \dev{y}\dev{x} \rangle$ of the positional correlations $C_{rr}$,
which are related to $f_{xy} \propto b$,
are expected to be small \cite{Oberlack:2006.1,Holzer:2009.2}.
For this reason $f_{xy}$ is neglected in the following.  
Both positional cross-correlations  $\langle \dev{x}\dev{y} \rangle$ and 
$\langle \dev{y}\dev{x} \rangle$
are proportional to the 
local shear rate $b$ and to the fluctuation strength $f_{yy}$ in the $y$-direction,
cf. Eq.~(\ref{E:shear_rxry}).

The autocorrelation of the velocity fluctuations $\langle \dev{v}_x^2 \rangle$ 
in the flow direction includes several
contributions. It depends  on the shear rate, the
second derivative of the flow and the fluctuation strength $f_{xx}$ and $f_{yy}$.
The cross-correlations between velocity and position appear only in the shear plane and they are proportional to the local shear rate.

The sub-matrices $\mat{C}_{vr}$ and $\mat{C}_{rv}$, describing the
cross-correlations between the positional and the velocity fluctuations, are related to the mean angular momentum of the particle:
\begin{align}
        \langle (m \vec{r}\times\vec{v})_{z} \rangle = m\langle \dev{x}
\dev{v}_y - \dev{y} \dev{v}_x \rangle 
	= - m b \langle \dev{y}^2 \rangle\,.
\end{align}

\subsection{Distribution of position and velocity}
\label{S:distribution}

Integrating the particle distribution function $\mathcal{P}(\dev{\vec{q}})$
in Eq.~\eqref{E:gauss_dist} with respect to its velocity degrees
of freedom one obtains the
particle's positional distribution function $\mathcal{P}(\dev{\vec{r}})$, which
may be expressed in terms of the covariance matrix 
$\mat{C}_{rr}$ as follows:
\begin{align}
\mathcal{P}(\dev{\vec{r}}) &= \int d\dev{\vec{v}}\,\mathcal{P}(\dev{\vec{q}}) \notag \\
		     &=\frac{1}{\sqrt{(2\pi)^3 \det(\mat{C}_{rr})}} 
\exp\left(
	-\frac{1}{2}\dev{\vec{r}}^T \mat{C}^{-1}_{rr} \dev{\vec{r}}
     \right)\,.
\label{E:gauss_r}
\end{align}

\begin{figure}[ht]
         \centering
         \includegraphics{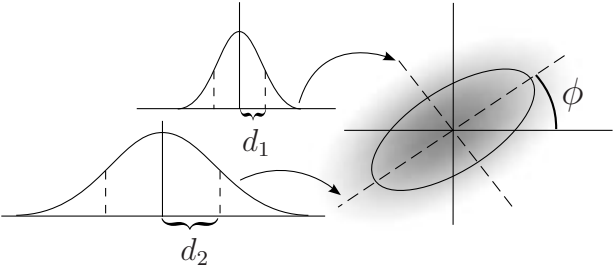}
         \caption{The elliptical particle distribution in the $xy$ shear plane
is sketched for a non-zero shear rate. $d_1$ and $d_2$ are the two principal axes of 
an ellipse along which $\mathcal{P}(\dev{\vec{r}})$ is constant. The Gaussian profiles
along the two principal axes are also indicated.}\label{F:winkel_rechts}
\end{figure}

A sketch of the distribution $\mathcal{P}(\dev{\vec{r}})$ is shown in
Fig.~\ref{F:winkel_rechts} for a linear shear flow $b \not =0$ and
$a=c=0$. 
The non-diagonal elements of the symmetric matrix $C_{rr}$, cf. Eq.~\eqref{E:crr} and Eq.~\eqref{E:shearall}, describe the
cross-correlations of the particle fluctuations in the $x$- and $y$-direction and therefore
the deviation of $\mathcal{P}(\dev{\vec{r}})$ from a spherically symmetric
distribution to an ellipsoidal one in the shear plane.
The principal axes of the particle's positional distribution are given by the
eigenvectors $\vec{w}_{p;1,2,3}$ of the corresponding eigenvalues $c_{p;1,2,3}$
of the matrix $\mat{C}_{rr}$.

In the $z$-direction one has $c_{p;3} = \langle \tilde{z}^2 \rangle$ and $\vec{w}_{p;3} =
(0,0,1)$. For the other two directions in the $xy$ shear plane
one obtains
\begin{align}
c_{p;1,2} &= \frac{1}{2}\left( \langle \dev{x}^2 \rangle
                                + \langle \dev{y}^2 \rangle
                      \right) \label{E:eigenvals} \\
           &\quad \pm \frac{1}{2} \sqrt{ 
	   		4 \langle \dev{x}\dev{y} \rangle^2
                      + \left(  \langle \dev{x}^2 \rangle - \langle \dev{y}^2
\rangle \right)^2
                   }, \notag \\
\vec{w}_{p;1} &= 
		  \left(
                                \frac{c_{p;1}-\langle \dev{y}^2\rangle}
				     {\langle \dev{x}\dev{y} \rangle} ,
                                1,
				0
                        \right), \notag \\
\vec{w}_{p;2} &= 
			\frac{\langle \dev{x} \dev{y}\rangle}
			     {|\langle \dev{x} \dev{y} \rangle|}
                        \left( -1,
                               \frac{c_{p;1}-\langle \dev{y}^2 \rangle}
				    {\langle \dev{x}\dev{y} \rangle},
				0
                        \right) \notag \,.
\end{align}
With the orthogonal transformation matrix
\begin{align}
        \mat{D} &= (\vec{w}_{p;1}, \vec{w}_{p;2}, \vec{w}_{p;3}), \label{E_matD}
\end{align}
the $\dev{\vec{r}}$-dependence of the exponential function
\eqref{E:gauss_r} can be rewritten as follows
\begin{align}
        -\frac{1}{2}\dev{\vec{r}}^T~ \mat{C}_{rr}^{-1}~ \dev{\vec{r}}
        = -\frac{1}{2} \left( \mat{D}^T \dev{\vec{r}} \right)^T
                \text{diag}\left(c_{p;1}^{-1},c_{p;2}^{-1}, c_{p;3}^{-1}\right)
                \left( \mat{D}^T \dev{\vec{r}} \right) 
\,. \label{E:exp_argument}
\end{align}
The eigenvalues $c_{p;1}$ and $c_{p;2}$ determine the length of the principal axes
$d_{p;1,2} = \sqrt{c_{p;1,2}}$ along the directions $\vec{w}_{p;1}$ and
$\vec{w}_{p;2}$ of an ellipse, where the longer one is
rotated counterclockwise with respect to the $x$-axis by an angle $\phi_p$, which
is given by the expression
\begin{align}
        &\tan \phi_p = 
	 \frac{\langle \dev{x}\dev{y} \rangle}
              {c_{p,1}-\langle \dev{y}^2\rangle} \label{E:tanphi} \\
		&= \frac{1}{2} \Bigg[\frac{\langle \dev{x}\dev{y} \rangle}{|\langle \dev{x}\dev{y} \rangle|}
				\sqrt{  4 + \left(\frac{\langle \dev{x}^2 \rangle - \langle \dev{y}^2 \rangle}
						    {\langle \dev{x}\dev{y} \rangle}
					    \right)^2
				}  
				- \left( \frac{\langle \dev{x}^2 \rangle - \langle \dev{y}^2 \rangle}
				              {\langle \dev{x}\dev{y} \rangle}
				  \right)
			\Bigg]\,. \notag		
\end{align}
The ratio between the minor and the major axis is given by
\begin{align}
        V_p :&= \sqrt{\frac{c_{p;2}}{c_{p;1}}} \label{E:pos_ratio} \\
	     &= \left( \frac{ \langle \dev{x}^2 \rangle + \langle \dev{y}^2 \rangle - \sqrt{ 4\langle \dev{x}\dev{y} \rangle^2 + (\langle \dev{x}^2 \rangle - \langle \dev{y}^2 \rangle)^2 }}
                    { \langle \dev{x}^2 \rangle + \langle \dev{y}^2 \rangle + \sqrt{ 4\langle \dev{x}\dev{y} \rangle^2 + (\langle \dev{x}^2 \rangle - \langle \dev{y}^2 \rangle)^2 }} \right)^{\frac{1}{2}} \notag \,.		
\end{align}

Analogous to the particle distribution given above, 
one can also derive an expression for the probability 
distribution $\mathcal{P}(\dev{\vec{v}})$ of the particle's velocity, which is
obtained by integrating out the positional degrees of freedom in Eq.~\eqref{E:gauss_dist}:
\begin{align}
\mathcal{P}(\tilde{\vec{v}}) &= \int{d\tilde{\vec{r}} \mathcal{P}(\tilde{\vec{r}},\tilde{\vec{v}})}\notag\\
		     &= \frac{1}{\sqrt{(2\pi)^3 \det(\mat{C}_{vv})}} 
	\exp\left(-\frac{1}{2}\dev{\vec{v}}^T \mat{C}_{vv}^{-1} \dev{\vec{v}}
	    \right)\,. \label{E:gauss_v}
\end{align}
The eigenvalues $c_{v;1,2,3}$ and eigenvectors $\vec{w}_{v;1,2,3}$  
of $C_{vv}$ are determined in a similar manner as for $C_{rr}$. Again the eigenvalue $c_{v;3}=\langle \tilde{v}_z^2 \rangle$ and the principal axis $\vec{w}_{v;3}=(0,0,1)$
perpendicular to the shear plane are obvious. The remaining two eigenvalues
and eigenvectors are:
\begin{align}
c_{v;1,2} &= \frac{1}{2}\left( \langle \dev{v}_x^2 \rangle
                                + \langle \dev{v}_y^2 \rangle
                        \right) \notag \\
           &\quad \pm \frac{1}{2} \sqrt{ 
	   		4 \langle \dev{v}_x \dev{v}_y \rangle^2
                      + \left(  \langle \dev{v}_x^2 \rangle - \langle
\dev{v}_y^2 \rangle \right)^2
                   }, \label{E:eigenvals_v} \\
\vec{w}_{v;1} &= 
		  \left(
                                \frac{c_{v;1}-\langle \dev{v}_y^2\rangle}
				     {\langle \dev{v}_x \dev{v}_y \rangle} ,
                                1,0
                        \right), \notag \\
\vec{w}_{v;2} &= 
			\frac{\langle \dev{v}_x \dev{v}_y\rangle}
			     {|\langle \dev{v}_x \dev{v}_y \rangle|}
                        \left( -1,
                                \frac{c_{v;1}-\langle \dev{v}_y^2\rangle}
				     {\langle \dev{v}_x \dev{v}_y \rangle},0
                        \right)
\, .
\end{align}
They have the same structure like those for
the covariance matrix $C_{rr}$. However, while the non-diagonal elements of $C_{rr}$
are directly proportional to the local shear rate $b$, the non-diagonal elements of
$C_{vv}$ and therefore the angle enclosed between
 the  principal axis of the distribution of the
velocity fluctuations and the $x$-axis, 
\begin{align}
        &\tan \phi_v = 
	 \frac{\langle \dev{v}_x\dev{v}_y \rangle}
              {c_{v,1}-\langle \dev{v}_y^2\rangle} \label{E:tanphi_v} \\
		&= \frac{1}{2}\Bigg[\frac{\langle \dev{v}_x\dev{v}_y \rangle}{|\langle \dev{v}_x\dev{v}_y \rangle|}
				\sqrt{  4 + \left(\frac{\langle \dev{v}_x^2 \rangle - \langle \dev{v}_y^2 \rangle}
						    {\langle \dev{v}_x\dev{v}_y \rangle}
					    \right)^2
				}  
				- \left( \frac{\langle \dev{v}_x^2 \rangle - \langle \dev{v}_y^2 \rangle}
				              {\langle \dev{v}_x\dev{v}_y \rangle}
				  \right)
			\Bigg]\,, \notag	
\end{align}
depends only weakly on the local shear rate, since the contribution of $b$ to the shear-induced cross-correlation $f_{xy}$ is small in magnitude, compared to $f_{ii}$ \cite{Oberlack:2006.1,Holzer:2009.2}.
From a measurement  of the distribution of the particle's velocity
one may calculate the eigenvalues  $c_{v;1,2}$ and the
angle $\phi_v$, which may enable the determination of the stochastic force correlations
$f_{xx}$, $f_{yy}$ and $f_{xy}$.

The mean kinetic energy of a trapped particle in shear flow 
is composed of two contributions, one induced by the distribution of the fluctuations and an
additional one by virtue of the rotational part of the flow:
\begin{align}
	E_{kin} &= \frac{m}{2} \langle \dev{v}_x^2 + \dev{v}_y^2  + \dev{v}_z^2
\rangle \label{E:ekin} \\
		&=   \frac{1}{4}\frac{f_{xx} + f_{yy} + f_{zz}}{m^2\beta} 
		+ \frac{b^2}{2}\langle \dev{y}^2 \rangle
		+
		\frac{8}{3} c^2 \frac{\tau_p^2 \omega^2 }
		{1+2\tau_p^2 \omega^2}\langle \dev{y}^2
\rangle^2\,. \notag
\end{align}
If the correlations of the stochastic forces are assumed to be 
independent of the flow, the mean energy of the particle will be increased
by the flow, since all terms in this equation are positive.
But without knowing the explicit expressions for the force correlations, 
it is not clear how the energy is really changed.
Recent calculations show 
that the mean kinetic energy of a fluid 
without an immersed particle increases in a shear flow \cite{Bamieh:2001,zarate:2008}. This indicates that the 
stochastic forces and the particle's mean energy, Eq.~\eqref{E:ekin}, may be 
amplified as well.

\section{Results for specific flows}
\label{S:results}

The results concerning the effects of a linear shear 
and a plane Poiseuille flow on the Brownian motion of a
particle in a harmonic potential are presented in this section. 
 Namely the properties of
the correlation functions and the particle's distribution are analyzed 
as a function of the flow parameters. 
They share common trends for both flows but there are also some
 characteristic differences, which are described in this section.

\subsection{Linear shear flow}
\label{S:shear}

If the center of the harmonic potential at $ \vec{r}_0=(0,y_0,0)$ does not coincide with the center of the linear shear flow of shear rate $\dot{\gamma}$, we may insert in Eq.~(\ref{E:flow}) $y = y_0 + \tilde y$, where $\tilde y$ describes the deviation from $y_0$. 
Consequently, in terms of the coordinates $\tilde y$, the three coefficients in Eq.~\eqref{E:flow} take the following form
\begin{align}
\label{E:abc_pois}
b&=\dot{\gamma}\,,&a&=\bar{a}=\dot{\gamma} {y}_0
&\text{and}\,\,\,\,c&=0\,.
\end{align}
In the case $y_0 \neq 0$ the flow velocity has a finite value at the center of the trap resulting in a non-zero mean position of the particle in flow direction, which is given via Eq.~\eqref{E:mean_x} by the formula
\begin{align}
 	\langle x \rangle &= \dot{\gamma}\tau_{p}~y_0\,.\label{E:shear_mean}
\end{align}

With the identifications \eqref{E:abc_pois} 
the elements of the covariance matrix $\mat{C}$  for the 
particle fluctuations around the potential minimum may be determined
for a linear shear flow by Eqs.~\eqref{E:shearall}
in terms of the noise amplitudes $f_{xx}$, $f_{xy}$
and $f_{yy}$.

An interesting question is, 
how to detect the noise amplitudes  $f_{ij}$
 in terms of the measured auto- and cross-correlations of the
particle's position- and velocity fluctuations. According to 
Eq.~(\ref{E:vxvy}) there is a direct relation
between the velocity correlation 
$\langle \tilde v_x \tilde v_y \rangle$ and the noise magnitude $f_{xy}$. 
Therefore, a direct measurement of this velocity correlation, if
experimentally possible, would give $f_{xy}$, or,
the other way around, a non-zero value of $f_{xy}\not =0$ is required
in order to obtain non-zero values of the
cross-correlations between these orthogonal velocity components.

As already mentioned in the introduction,
it is favorable to investigate 
the particle fluctuations around a
potential minimum rather than those of free particles since trapped particles can be investigated 
over a long period of time, as
demonstrated by several experiments, see e. g.  Ref.~\cite{Quake:99.1}.
This opens the opportunity for the determination of the
magnitudes of the noise $f_{ij}$ in terms of the positional
fluctuations via Eqs.~(\ref{E:shearall}),
which in the case of a linear shear flow take the following
explicit form 
\begin{subequations}
\begin{align}
	f_{yy} &= 2 k \zeta \langle \dev{y}^2 \rangle
	\,, \\
	f_{xy} &= 2 k \zeta \Bigg(
		  \langle \dev{x}\dev{y} \rangle
		- \frac{1}{2}\dot{\gamma} \tau_p
			\langle \dev{y}^2 \rangle
		  \Bigg)
	\,,  \\
	f_{xx} &= 2 k \zeta \Bigg(
			\langle \dev{x}^2 \rangle
			- \frac{1}{2} \frac{\dot{\gamma}^2}{\omega^2} 
			\langle \dev{y}^2 \rangle
			- \dot{\gamma} \tau_p \langle \dev{x}\dev{y} \rangle
		\Bigg)\,.
\end{align}
 \label{E:shear_f}
\end{subequations}
In the case of isotropic stochastic
forces with negligible cross-correlations,  
$f_{xx}=f_{yy}$ and $f_{xy} \propto \langle \tilde v_x \tilde v_y \rangle \approx 0$,
one has in a linear shear flow still 
a non-vanishing positional cross-correlation $\langle \tilde x \tilde y \rangle
= \dot \gamma \tau_p \langle \tilde y^2 \rangle/2$. Its
magnitude is determined by the shear rate $\dot \gamma$ and
the noise strength via $\langle \tilde y^2 \rangle$. According to
this behavior and because $\langle \tilde x^2 \rangle \not = \langle \tilde y^2 \rangle$,
we expect an anisotropic distribution of the
positional fluctuations $\cal{P}(\dev{\vec{r}})$, as sketched  
in Fig.~\ref{F:winkel_rechts} and as discussed below. For $f_{xy} \neq 0$ 
the anisotropy of the positional distribution has an additional contribution
that depends on the magnitude of $f_{xy}$.

During the rest of the present section~\ref{S:shear} inertia effects are neglected and
in addition, we 
assume an isotropic 
noise distribution with $f_{xy}=f_{yx}=0$ and $f_{xx}=f_{yy}=f_{zz}= 2 k_B T \zeta$.
Both are good approximations for many experiments 
focusing on leading order effects of a 
shear flow on the fluctuations of particles. 
Taking into account the time-dependence
of the positional fluctuations, as given by Eqs.~(\ref{E:sol_r}), 
one obtains in terms of the dimensionless  Weissenberg number, 
\begin{align}
 \Wi = \tau_p \dot \gamma \,,
\end{align}
and the Heaviside step function $\Theta(t)$ 
the following time-dependent correlations:
\begin{subequations}
\begin{align}
\langle \dev{x}(t)\dev{x}(0) \rangle
        &=  \frac{k_BT}{k} \left( 1  + \frac{\Wi^2}{2}~\bigg(1+\frac{|t|}{\tau_p}\bigg) \right) \displaystyle e^{-|t|/\tau_p}
	   \,, \label{E:corr_shear} \\
\langle \dev{y}(t)\dev{y}(0) \rangle
        &=  \langle \dev{z}(t)\dev{z}(0) \rangle=
\frac{k_BT}{k} e^{- |t|/\tau_p}
	   \,, \\
\langle \dev{x}(t)\dev{y}(0) \rangle
        &=   \frac{k_BT}{k}~\frac{\Wi}{2} \left( 1  + 2\frac{t}{\tau_p}~ \Theta(t) \right) e^{- |t|/\tau_p}
	   \,. 
\label{E:crosscorrxy}
\end{align}
\label{E:crosstime}
\end{subequations}
The cross-correlation between fluctuations along orthogonal directions,
as given by the last equation,  
is shear-induced and its asymmetry 
$\langle \dev{x}(t)\dev{y}(0) \rangle \not = \langle \dev{x}(0)\dev{y}(t) \rangle$
with respect to time reflection $t \to -t$ is one of the important effects
of  shear flow on the distribution of fluctuations.

\begin{figure}[ht]
   \begin{center}
   \includegraphics{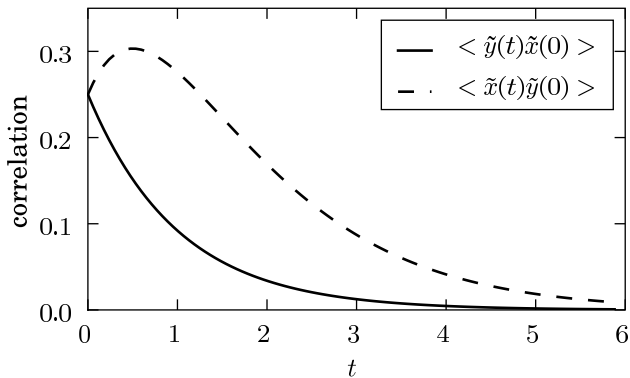}
   \end{center}
   \caption{
Time dependence of the shear-induced cross-correlations
$\langle \tilde x(t) \tilde y(0) \rangle$ (dashed line) 
and $\langle \tilde x(0) \tilde y(t) \rangle$ (solid line)
 as given by Eq.~\eqref{E:crosscorrxy} for the
  Weissenberg number $\Wi=1$ and the relaxation time $\tau_p=1$.
}
 \label{F:time_shear}
\end{figure}

For $t>0$, the algebraic prefactor in Eq.~(\ref{E:crosscorrxy}) illustrates that 
a fluctuation $\tilde y(0)\not =0$ of a particle is carried away
by the flow along the $x$-direction before the initial 
displacement starts to relax remarkably. This leads,
during an initial period of time shorter than the relaxation time $\tau_p$,
to the growth of $\langle \tilde x(t) \tilde y(0) \rangle $, 
while $\langle \dev{x}(0) \dev{y}(t) \rangle $
decays monotonically (solid line in Fig.~\ref{F:time_shear} corresponding
to $t<0$). 
The predicted elementary signatures for
the shear-induced cross-correlations, as shown in Fig.~\ref{F:time_shear},
are in agreement with
 experimental data as described recently in Ref.~\cite{Ziehl:2009.1}.
The expression $\langle \dev{x}(t) \dev{y}(0) \rangle $ is proportional to the 
Weissenberg number and takes its maximum at half of the  particle's 
relaxation time $t_{max} = \tau_p/2$.
 For the correlations of the velocity fluctuations of the fluid in orthogonal directions 
 a similar signature as in Eq.~(\ref{E:crosscorrxy}) has been found
 \cite{Eckhardt:2003.1}, where however the 
mechanism is slightly different.

A comparison of the absolute values of the particle's fluctuations
with experimental results may be difficult. However, one obtains
from Eqs.~(\ref{E:crosstime}) and in terms of
the dimensionless Weissenberg number the following 
normalized ratios of the static correlations
\begin{subequations}
\label{E:corr_ratio}
\begin{align}
& \displaystyle 
\frac{ \langle \tilde x(0) \tilde y(0) \rangle}{  \langle \tilde y(0) \tilde y(0) \rangle} = \frac{\Wi}{2} \,, \\
& \frac{\langle \tilde x(0) \tilde y(0) \rangle}{\langle \tilde x(0) \tilde x(0) \rangle} = \frac{\Wi/2}{1+\Wi^2/2}\,.
\end{align}
\end{subequations}
The left hand side and the right hand side of Eqs.~\eqref{E:corr_ratio} can be measured independently in different experiments and the results can be compared afterwards.

\begin{figure}[ht]
    \begin{center}
   \includegraphics{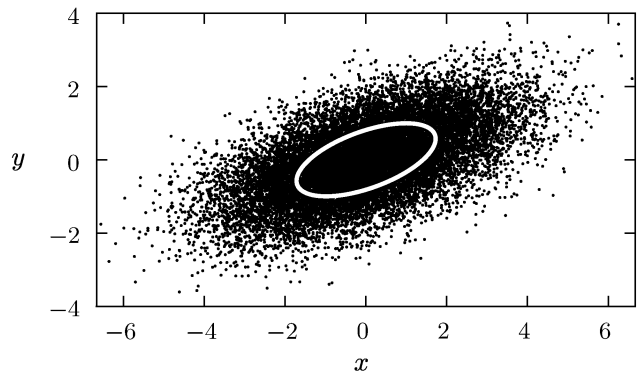}
   \end{center}
    \caption{Positional distribution of a particle in a harmonic
potential with its minimum at ${\bf r}_0=(0,0,0)$ and exposed to
a linear shear flow, as
 obtained by a stochastic simulation 
of the Eqs.~\eqref{E:lang_equation} in the over-damped limit 
for the Weissenberg number  $\Wi=2$ and the relaxation time $\tau_p=1$.}
 \label{F:dist_shear}
\end{figure}

An anisotropic distribution of the particle's velocity 
$\cal{P}(\tilde {\bf v})$, as given by Eq.~(\ref{E:gauss_v}),
is only obtained in the case of a finite 
cross-correlation $f_{xy}$ of the stochastic
forces. In contrast to this, 
the particle's distribution   $\cal{P}(\tilde{\bf r})$
has in the overdamped
limit an elliptical shape in the $xy$ plane, even for a vanishing
cross-correlation magnitude $f_{xy}=0$.
In this limit an
 elliptical distribution $\cal{P}(\tilde{\bf r})$
is shown in Fig.~\ref{F:dist_shear} for the Weissenberg 
number $\text{Wi}=2$ and the relaxation time $\tau_p=1$. 
In Fig.~\ref{F:dist_shear} the time evolution 
 of the particle's position, obtained by 
simulations of the basic equation (\ref{E:lang_equation}),
is plotted at equidistant times. 
This positional distribution can be characterized by 
the angle $\phi_p$ of the ellipsoid and the ratio $V_p$ of its principal axis. The
 general expressions for $\phi_p$ and $V_p$, as given by Eqs.~\eqref{E:tanphi} and \eqref{E:pos_ratio},
 can be further
simplified in the case of  a linear shear flow to
 functions of the dimensionless 
Weissenberg number $\Wi$ only:
\begin{subequations}
\label{E:tanp_V_s}
\begin{eqnarray}
 \tan \phi_p &=& \frac{1}{2}
\left[ \sqrt{4+\Wi^2}  - \Wi \right]\,,
\label{tanp}\\
 V_p&=&\left(\frac{\sqrt{4+\Wi^2}-\Wi}{\sqrt{4+\Wi^2}+\Wi}\right)^{1/2}\,.
\label{V}
\end{eqnarray}
\end{subequations}

The two expressions in Eqs. \eqref{E:tanp_V_s}
 suggest measurements  
of the shear-induced particle fluctuation effects, which are
complementary to the measured static correlations.
In experiments the particle positions may be recorded at equidistant times.
By plotting these subsequent particle positions in the shear plane,  a similar 
distribution is expected as shown by our numerical simulation
in Fig.~\ref{F:dist_shear}. From such an experimentally 
measured distribution for different shear rates
the angle $\phi_p$ and the ratio between the principal axes, $V_p$, may
be determined.
If the determination of the Weissenberg number $\Wi$ is difficult or if the precision is not sufficient Eqs. \eqref{E:corr_ratio} and \eqref{E:tanp_V_s} allow a consistency check between different aspects of the particle fluctuations, without a separate measurement of $\Wi$.
A cross-check has recently been performed in an experiment  
in which a good agreement between both approaches
has been found, cf. Ref.~\cite{Ziehl:2009.1}.

\begin{figure}[ht]
   \begin{center}
   \includegraphics{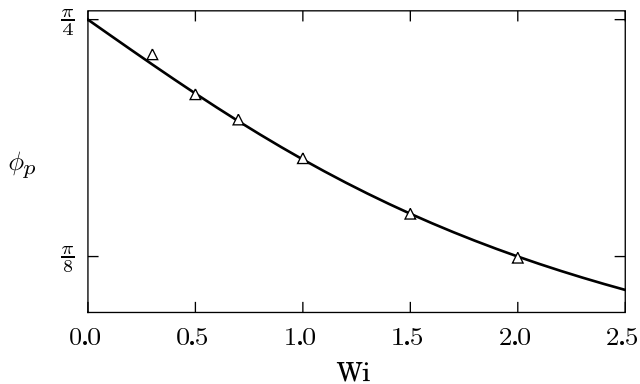}
   \end{center}
   \caption{
The angle of the ellipse in the $xy$ plane with respect to the flow lines as a function of the 
Weissenberg number $\Wi$, as it is given by Eq.~\eqref{tanp}. The triangles are the result of a 
 simulation of Eq.~\eqref{E:lang_equation} in the over-damped limit.
}
 \label{F:phi_shear}
\end{figure}
\begin{figure}[ht]
   \begin{center}
   \includegraphics{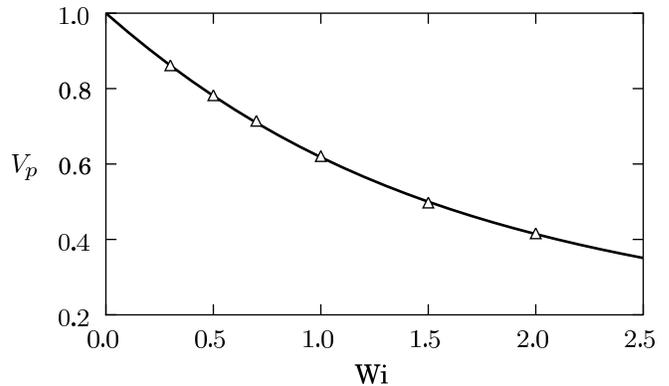}
   \end{center}
  \caption{The ratio $V_p$ of the principal axis of the ellipse in Eq.~\eqref{V}, as a function of the Weissenberg number $\Wi$. 
The triangles are the result of a simulation of Eq.~\eqref{E:lang_equation} in the over-damped limit.
}
 \label{F:ratio_shear}
\end{figure}

In the limit of a vanishing Weissenberg number, $\Wi \to 0$,
the angle $\phi_p$ of the ellipsoidal 
particle's distribution tends to $\phi_p=\pi/4$ and the positional variance
 becomes isotropic, $\langle \dev{x}^2 \rangle = \langle \dev{y}^2 \rangle$,
 corresponding to
the ratio $V_p=1$. 
This trend is similar to the dependence of the orientation of
vesicles in shear flow \cite{Kraus:96} or to the 
local orientation of the order parameter of nematic liquid crystals 
in plane shear flows \cite{deGennes:2006}.

If $\dot{\gamma}$ and therefore $\Wi$ 
 is increased, the angle $\phi_p$ between the longer
 semi-axis and the $x$-axis as well as the ratio $V_p$ 
will decrease as shown in Fig.~\ref{F:phi_shear} and in Fig.~\ref{F:ratio_shear}.
The triangles in those figures
have been obtained by numerical simulations of Eq.~(\ref{E:lang_equation})
for the same values of the Weissenberg number as used for the
analytical curves. In our simulations of the Langevin model
 we have assumed an isotropic and Gaussian
distributed white noise and vanishing cross-correlations $f_{xy}=0$.
The simulation results confirm the assumption of a Gaussian distribution 
of the particle positions in the case of a shear flow in Sec.~\ref{S:fluct}.
The analytical and numerical results on $\phi_p$ and $V_p$
in Fig.~\ref{F:phi_shear} and in Fig.~\ref{F:ratio_shear} are in
 good agreement.

\subsection{Plane Poiseuille flow}
\label{S:poiseuille}

A Brownian particle is trapped by a harmonic potential
 close to $\vec{r}_0=(0,y_0,0)$ and
exposed to a plane Poiseuille flow along the $x$-direction,
\begin{align}
	\vec{u}(y)= u_p\left(1-\frac{y^2}{l^2}\right)\ex\,. 
	\label{E:pois_flow}
\end{align}
The particle fluctuations $\tilde x$ and $\tilde y$ describe 
deviations with respect to the mean values  $y_0$
and $\langle x \rangle$, which is not zero as determined below. 
The flow velocity along the $x$-direction 
may be expressed for further analysis in terms
 of the fluctuations $\tilde y$ as follows
\begin{align}
	u_{x}(\tilde y)= u_p\left(1-\frac{y_0^2}{l^2}\right) 
			- 2 \frac{u_p y_0}{l^2}  \tilde y 
				- \frac{u_p}{l^2} \tilde y^2\,.
\label{E:pois_flow_pot}
\end{align}
Comparing this expression with Eq.~\eqref{E:flow} the coefficients 
in the latter equation are given by
\begin{align}
c&=-\chi := -\frac{u_p}{l^2}\,,\notag \\
a&=u_p-\chi y_0^2  \,,\notag \\
b&= - 2 \chi y_0 \,. \label{E:coeff_poiseuille}
\end{align}
Here $\chi$ describes the second derivative of the velocity profile,  $b$ the local shear rate and 
\begin{align}
 \text{Wi}(y_0):= -2 \tau_p \chi y_0\,,
\end{align}
the local Weissenberg number $\text{Wi}(y_0)$.
With these identifications 
the elements of the covariance matrix $\mat{C}$  are again
given via the expressions in Eqs.~\eqref{E:shearall}
in terms of the strength of the noise,  $f_{xx}$, $f_{xy}$, $f_{yy}$
and $f_{zz}$ and the flow parameters.

The mean position of the particle in a plane Poiseuille flow
can be determined via Eq.~\eqref{E:mean_x}:
\begin{align}
	\langle x \rangle &= \tau_p \left[ u_p  - \chi 
		\left(y_0^2 +\frac{1}{3}R^2\right)
	- \chi  \langle \tilde y^2 \rangle\, \right].\label{E:pois_mean}
\end{align}
In contrast to a linear shear flow, it includes  
a contribution depending on the  particle's radius $R$, which
is a pure deterministic effect due to Fax\'en's theorem \cite{Faxen:27.1,Dhont:96}.
The last term on the right hand side describes an additional shift based on the 
positional variance $\langle \tilde y^2 \rangle$ in the $y$-direction. 
Both contributions are proportional to the second derivative $\chi$ of the
flow profile and are therefore not present in linear shear flows.

In experiments Eq.~\eqref{E:pois_mean}
may be used to measure the
spatial variation of the flow profile by detecting the 
mean displacement of a particle of radius $R$
out of the optical tweezer potential.
For such a measurement usually deterministic formulas are used to describe the relation between
the displacement and the flow velocity. But Eq.~\eqref{E:pois_mean}
indicates that a correction due to
thermal motion has to be taken into account.

The relations given by Eqs.~\eqref{E:shearall} relate 
the fluctuations of the velocity and the position of a particle to the externally controlled flow properties
and the magnitudes of the thermal fluctuations. Consequently they allow, in a similar manner as to the linear shear flow, a determination of
the magnitudes of the stochastic forces in terms of the measured covariances 
of the particle fluctuations:
\begin{subequations}
\begin{align}
	f_{yy} &= 2 k \zeta \langle \dev{y}^2 \rangle 
	\,, \\
	f_{xy} &= 2 k \zeta \Big(
		\langle \dev{x}\dev{y} \rangle
		-\frac{1}{2} \Wi(y_0)~
		\langle \dev{y}^2 \rangle \Big)
	\,,  \\
	f_{xx} &= 2 k \zeta \Bigg(
			\langle \dev{x}^2 \rangle
			- \frac{1}{2}  \frac{\Wi(y_0)}{\tau_p \omega^2}
			\langle \dev{y}^2\rangle \notag \\
		&
		- \frac{2}{3} \tau_p^2 \chi^2 \Bigg(
		\frac{ 5 + 2\tau_p^2 \omega^2 }
		     { 1 + 2\tau_p^2 \omega^2 }
		\Bigg)\langle \dev{y}^2 \rangle^2
		- 
		\Wi(y_0) \langle \dev{x}\dev{y} \rangle 
		\Bigg)\,.
\end{align}
\label{E:fii_pois}
\end{subequations}
The difference compared to Eqs.~\eqref{E:shear_f} is an additional contribution to $f_{xx}$, 
which depends on the second derivative of the Poiseuille flow. Further comments made above
for a linear shear flow hold as well.

\begin{figure*}[ht]
    \begin{center}
   \includegraphics{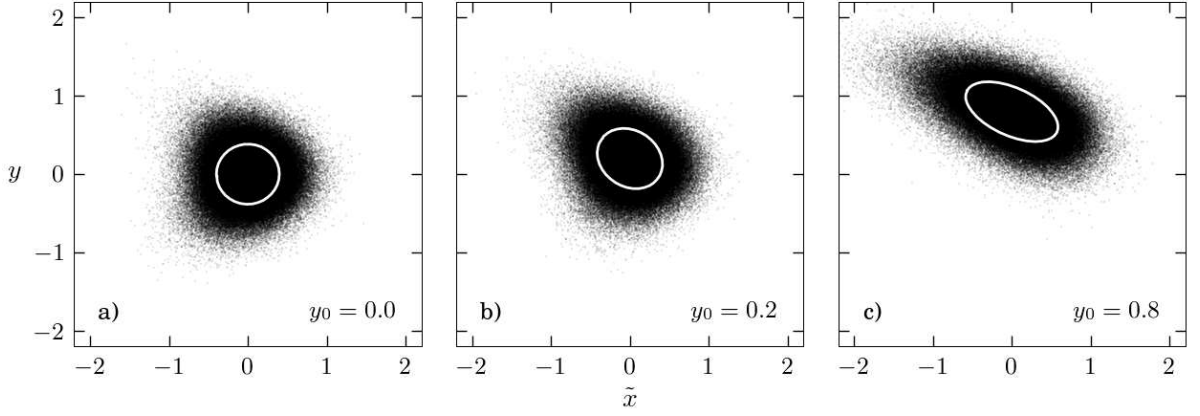}
   \end{center}
    \caption{Positional distribution of a Brownian particle in the $xy$ plane
captured by a harmonic
potential with its minimum at $x_0=z_0=0$ 
and for different values of $y_0$. The particle is  exposed to
a Poiseuille flow and its dynamics have been calculated by stochastic simulations 
of the Eqs.~\eqref{E:lang_equation} in the over-damped limit 
for $\chi=1$, $k_BT=0.1$ and $\tau_p=1$.}
 \label{F:dist_pois}
\end{figure*}

Similar to the end of the previous section \ref{S:shear},
we neglect from here on the particle inertia and we assume in addition isotropically distributed 
noise with $f_{xy}=0$ and $f_{ii}=2k_BT\zeta$. The
static correlations of the particle's positional fluctuations in the shear plane,
given by Eqs.~(\ref{E:shearall}), 
then reduce to
\begin{subequations} \label{E:pois_corr_m0}
\begin{align}
\langle \dev{x}^2 \rangle
        &=   \frac{k_BT}{k} \left(1  
	   +  \frac{1}{2} \Wi(y_0)^2
	       + \frac{2}{3} \chi^2 \tau_p^2 \frac{k_BT}{k} \right)
	   \,, \label{E:pois_rx2_m0} \\
\langle \dev{y}^2 \rangle&=\frac{k_BT}{k}\,,\label{E:pois_ry2_m0} \\
\langle \dev{x}\dev{y} \rangle &= \langle \dev{y}\dev{x} \rangle 
        =    \frac{k_BT}{k}  \frac{\Wi(y_0)}{2}
	  \,. \label{E:pois_rxry_m0}
\end{align}
\end{subequations}
The static cross-correlation $\langle \dev{x}\dev{y} \rangle$ has the same 
dependence on the Weissenberg number as in the linear shear case Eq.~\eqref{E:crosscorrxy} at $t=0$,
if the local shear rate of the Poiseuille 
flow at the potential center is taken.
In contrast to a linear shear flow profile, the mean-square displacement of the particle in $x$-direction, given by Eq.~(\ref{E:pois_rx2_m0}),
includes an additional constant contribution, $G:=2/3\left( \chi \tau_p k_BT/k \right)^2$, which
is independent of the position of the potential minimum in the Poiseuille flow and therefore
also independent of the local Weissenberg number. Besides the dependence on the 
second derivative $\chi$ of the flow profile, the contribution $G$ is a direct function of  
 the fluid temperature. Due to this contribution in Eq.~\eqref{E:pois_rx2_m0} 
one has $\langle \tilde x^2 \rangle \neq \langle \tilde y^2 \rangle$
and 
the isotropy of the particle's positional distribution in
the shear plane 
is broken. This broken rotational symmetry also
changes the analytically determined distribution
function in Sec.~\ref{S:distribution} along with the
correlation matrix $\mat{C}_{ij}$ for a Poiseuille flow.

In addition to the calculations, we have performed simulations of the particle dynamics
where we used isotropic and Gaussian distributed white noise in the overdamped version
of  Eq.~(\ref{E:lang_equation}). The resulting distributions for the particle's position are shown in Fig.~\ref{F:dist_pois} for three different positions $y_0$ of the potential center.
With the potential minimum at the center of the flow ($y_0=0$) 
the numerical results show a broken mirror symmetry in $x$-direction of the particle's positional distribution, c.f. Fig.~\ref{F:dist_pois}a).
One recognizes a parachute shape that is similar to the well known conformation
of vesicles and red blood cells in the center of a Poiseuille flow \cite{Misbah:2009.1}. 
If one now makes the heuristic assumption of
a Gaussian distribution for the particle's position
in the case of a Poiseuille flow, as 
in Sec.~\eqref{S:fluct}, but with the correlation matrix
$\mat{C}_{ij}$ determined in terms of the Poiseuille-flow parameters,
one expects for the parameters in Fig.~\ref{F:dist_pois}a) an elliptical shape of the particle
distribution. Indeed, the ratio between the principal axis in Fig.~\ref{F:dist_pois}a) is slightly smaller than $1.0$.
But within this analytical 
approximation, the $\pm x$ symmetry is not broken,
which indicates the limitation of the heuristic approach.

Away from the center of the Poiseuille flow, for 
finite values of $y_0 \neq 0$, the $\pm y$ symmetry of
the particle's positional distribution is also broken, as shown
in Fig.~\ref{F:dist_pois}b) and in Fig.~\ref{F:dist_pois}c).
With increasing values of $y_0$ the local shear
rate acting on the particle increases as well as the
 local Weissenberg number $\Wi(y_0)$.
Consequently the cross-correlation $\langle \dev{x}\dev{y} \rangle$ in Eq.~(\ref{E:pois_rxry_m0})
becomes non-zero and the particle's positional
distribution in the $xy$ plane approaches, according to our
analytical results, an elliptic shape as indicated
by the ellipses in  Fig.~\ref{F:dist_pois}b) and Fig.~\ref{F:dist_pois}c).
Again the full numerical simulations show deviations
from the elliptical shape.

\begin{figure}
   \begin{center}
   \includegraphics{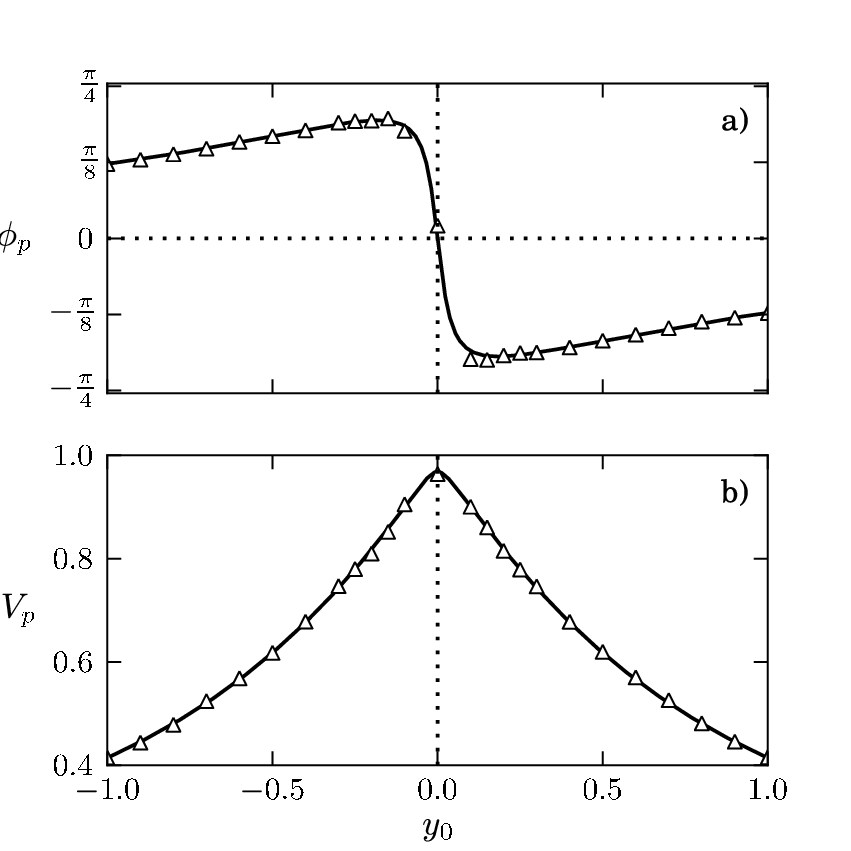}
   \end{center}
   \caption{The angle $\phi_p$ in the $xy$ plane according to Eq.~\eqref{E:tanphi_pois} is shown in a) and the ratio $V_p$ between the two principal axes, as 
determined by the Eqs.~\eqref{E:pois_corr_m0} and Eq.~\eqref{E:pos_ratio}, is shown in b). Both are plotted as a function of the position of the potential mini\-mum $y_0$ and for the parameters $\chi=\tau_p=1,\langle \dev{y}^2 \rangle=k_B T/k= 0.1 $.
The triangles  are the result of  simulations of the Eqs.~\eqref{E:lang_equation} with the same parameter set.
}
 \label{F:phi_ratio_pois}
\end{figure}

However, the inclination of the distribution and 
the inclination of the analytically
determined ellipses agree rather well and therefore
a determination  of the angle $\phi_p$ for a Poiseuille flow 
according to Eq.~(\ref{E:tanphi}), similar to that in the previous chapter, 
is reasonable and $\phi_p$ has the following form
\begin{align}
 \tan \phi_p &= \frac{1}{2}  \Bigg[ -\frac{y_0}{|y_0|} \sqrt{4 + \left(\Wi(y_0) 
	+ \frac{4}{3} \frac{\tau_p^2 \chi^2}{\Wi(y_0)}  \langle \dev{y}^2 \rangle\right)^2}  \notag \\
& ~~~~~~~~~~~~~~ - \left(\Wi(y_0) 
	+ \frac{4}{3} \frac{\tau_p^2 \chi^2 }{\Wi(y_0)} \langle \dev{y}^2 \rangle\right) \Bigg]\,. \label{E:tanphi_pois}
\end{align}
We have shown in Sec.~\ref{S:shear}, how the angle $\phi_p$ and the
ratio $V_p$ depend on the Weissenberg number of the linear shear flow.
Since the local shear rate in a Poiseuille flow 
depends on the location $y_0$
of the minimum of the potential, one may plot 
$\phi_p$ and $V_p$ (which is calculated via Eq.~\eqref{E:pos_ratio}) as a function of $y_0$  
as shown in Fig.~\ref{F:phi_ratio_pois}.
Since the particle's distribution 
is anisotropic in the $xy$ plane even for $y_0=0$, see Fig.~\ref{F:dist_pois}a), the corresponding principal axes are always unequal. This behavior is reflected in Fig.~\ref{F:phi_ratio_pois}b) where the inequality $V_p<1$ holds for all values of $y_0$.
The dependence of the angle $\phi_p$ on the local shear rate in a Poiseuille flow differs from the case of a linear shear flow:
the function $\phi_p(y_0)$ is always well defined, even at the center of the flow, where it vanishes. This is one consequence of the asymmetry of the particle distribution.
With increasing values of $|y_0|$ and therefore
with increasing values of the local
shear rate, the angle $|\phi_p|$ increases as well until it reaches 
some maximum value $\phi_{max}$ at $y_{0;max}$, as given by the following equations: 
\begin{subequations} \label{E:pois_max}
 \begin{align}
   &y_{0;max} =  \pm \sqrt{\frac{\langle \dev{y}^2 \rangle}{3}}\,, \label{E:pois_y0max} \\
  &\phi_{max} = \mp \arctan\left(\sqrt{1 + \frac{4}{3}\tau_p^2\chi^2\langle \dev{y}^2\rangle} - 2\tau_p\chi \sqrt{\frac{\langle \dev{y}^2\rangle}{3}} \right)\,. \label{E:pois_phimax}
 \end{align}
\end{subequations}
$y_{0,max}$  depends only on $\langle \tilde y^2  \rangle$ and therefore
on the width of the particle's distribution in $y$-direction, which
is determined by the ratio between the thermal energy and the 
spring constant related to the harmonic potential acting on the
 particle. 
Increasing  $|y_0|$ beyond $|y_{0,max}|$ the local shear rate, $\Wi(y_0)$, increases too, but
$\phi_p$ starts to decrease; a similar behavior as seen in section \ref{S:shear}.  In the range $|y_0| > |y_{0,max}|$ the 
local shear dominates the curvature
effects more and more and the Poiseuille flow 
resembles a linear shear flow.
 In addition, the particle's distribution 
approaches an ellipse as obtained by the heuristic 
approximation. As mentioned above, the heuristic analytical approach
becomes exact in the case of a linear shear flow.

In Fig.~\ref{F:phi_ratio_pois} the analytical results on $\phi_p$ and
$V_p$ have been compared with our  numerical simulations. In spite of the
fact that we assumed for our analytical calculations
a Gaussian distribution of the positional fluctuations $ \tilde x$ and $\tilde y$, the results show surprisingly good agreement. The major reason for this good
agreement is that we calculated the
inertia tensor for the particle distribution, which includes only second order moments as
assumed for the Gaussian distribution.

\subsection{Pipe Flow}

For a flow profile in a pipe of radius $d$, cylindrical polar-coordinates
are appropriate. Similar to a plane Poiseuille flow the dependence of the velocity profile on the radial position $\rho$ is also quadratic.
Consequently, the $\rho$ dependence corresponds now
to the $y$ dependence in the case of the plane Poiseuille flow.
The velocity profile for the pipe flow, where the pipe axis coincides with the $x$ axis, is of the following form,
\begin{align}
	u_{x}=u_p\left(1-\frac{\rho^2}{d^2}\right)\,.
\end{align}
The results described above in section~\ref{S:poiseuille} 
apply qualitatively also for the pipe flow.

With the potential minimum located at $\rho_0$ 
the coefficients in the Eq.~\eqref{E:flow} are 
$a=u_p(1-\frac{\rho_0^2}{d^2})$, $c=-\chi := -\frac{u_p}{d^2}$ and $b= 2 \rho_0 c$.
Moreover, one has to consider $\bar{a} = a + \frac{2}{3} R^2 c$ in Eq.~\eqref{E:force_hydro}.
The major difference to the plane Poiseuille flow is the prefactor $\frac{2}{3}$ instead of $\frac{1}{3}$. This is a consequence of the $2D$ Laplacian instead of a $1D$ second derivative in Eq.\eqref{E:Faxen}.
The  mean-position of the particle
is in terms of the pipe flow of the following form
\begin{align}
	\langle x \rangle &= u_p \tau_p 
	- \chi \tau_p (y_0^2 + \frac{2}{3}R^2)
			- \chi \tau_p \langle y^2 \rangle
\,.\label{E:pois_shear_mean2}
\end{align}
Besides the factor $2$ in front of $R^2$ this expression corresponds to that in Eq.~\eqref{E:pois_mean}.

\section{Conclusion}
\label{conclusion}

In this work we have calculated analytically and numerically the
autocorrelations and cross\--correlations between different components
of the velocity and the positional fluctuations of a Brownian particle
in a harmonic potential, which is exposed to 
different shear flows.  
In addition,  
the particle's probability distribution in the 
harmonic potential  has been
determined as a function of the flow parameters.
By solving an appropriate Langevin model,
 cross-correlations between velocity and positional fluctuations
along orthogonal directions have been found and several suggestions
for experimental measurements are made.

Cross-correlations between orthogonal velocity
components occur only if there is already a  cross\--correlation 
between stochastic forces along orthogonal directions in
the related Langevin model.
On the other hand, we find cross-correlations 
between particle fluctuations along orthogonal
directions without cross\--correlations of orthogonal
stochastic force components and their magnitude increases with the dimensionless
shear rate, the Weissenberg number.

There are recent calculations on shear-induced
cross-correlations between orthogonal fluctuations
of freely floating particles \cite{Bedeaux:1995.1,Drossinos:05}. 
They have the same origin as 
those discussed in this work. However,  
while measurements of the shear-induced 
cross\--correlations
of freely moving particles may be difficult, 
shear-induced 
cross\--correlations for particles in a potential
  can be measured in a controlled manner. 
This insight
was the basis of recent successful measurements on the cross\--correlations
between particle fluctuations along orthogonal directions as described in Ref.~\cite{Ziehl:2009.1}.
In this experiment, harmonic potentials for  small latex spheres 
are induced by optical tweezers. The particle is
simultaneously exposed to a linear shear flow in 
a special flow cell and its Brownian motion is 
investigated directly. In the same experiment  and
in a forthcoming work, the cross\--correlations 
and anti-correlations between two particles captured in two neighboring potentials 
and exposed to shear flows are also investigated.

The measurements on the positional cross-correlation, 
$\langle x(t) y(0) \rangle $, presented in Ref.~\cite{Ziehl:2009.1},
exhibit a similar maximum as predicted by the
 expression  in Eq.~(\ref{E:crosscorrxy}) and as shown in Fig.~\ref{F:time_shear}.
This maximum is a typical signature of Brownian motion in shear flow and
it is found where the shear rate approximately equals half of
the particle's relaxation time. Also the elliptic shape
of the particle's distribution, $P({\bf r})$, as shown in Fig.~\ref{F:dist_shear},
 has been measured  in  Ref.~\cite{Ziehl:2009.1}.
The angle $\phi_p$, enclosed by the major axis of the 
distribution $P({\bf r})$, and the ratio $V_p$ between the
length of the two principal axes on the one hand and  the magnitude of the
static correlations on the other hand focus on  different aspects of the dynamics of
a particle in a potential exposed to
a shear flow. The interrelation between these different aspects
of the Brownian particle dynamics can be used for consistency
checks in experiments; such
cross-checks are described  Ref.~\cite{Ziehl:2009.1} 
for the case of a linear shear flow. They are also 
possible in experiments with Poiseuille flow as forthcoming measurements
may demonstrate. 

Shear-induced correlations of fluctuations of the fluid
velocity with respect to a linear shear flow 
were found theoretically  \cite{Eckhardt:2003.1,Oberlack:2006.1}. 
They are traced back to the non-normal property of the Navier-Stokes 
equation linearized around the linear shear profile \cite{Eckhardt:2003.1}
and these velocity fluctuations play an  important role in a shear flow
for its 
instability and the onset of turbulence. 
The cross-correlations between these velocity fluctuations exhibit
a similar extremum as given by Eq.~(\ref{E:crosscorrxy}), but
it is a slightly different mechanism 
leading to this similar behavior on the level of the correlation function. This interrelation may be
discussed in  more detail in future work.

Stochastic forces acting on a particle in a fluid are caused
by the velocity fluctuations of the fluid surrounding a
particle. In quiescent fluids the correlations for
velocity fluctuations are 
isotropically distributed.  For reasons of 
simplicity, this is quite often assumed in models
to investigate the Brownian motion of particles in shear flow. In that case, cross-correlations between
 stochastic forces along orthogonal directions 
 vanish in the Langevin equation of motion.
Shear induced cross-correlations between orthogonal components 
of fluctuations of the fluid velocity  are indeed small compared 
to the shear independent contributions
\cite{Oberlack:2006.1,Holzer:2009.2} and hence this simplification
is reasonable. However, to which extent such cross\--correlations
of the fluid velocity
may quantitatively  modify the presented results on the 
cross\--correlations of particle displacements 
needs further investigations.

In numerical simulations of the dynamics of 
a Brownian particle in a potential and exposed to a Poiseuille flow
we found higher order correlations for the 
particle fluctuations of non-Gaussian
behavior. However, for deriving our analytical
results on the
angle $\phi_p$, which the major axis of the
particle distribution encloses with
the flow direction, and the ratio $V_p$ between the lengths of the
two principal axes, we 
assumed Gaussian distributed 
particle fluctuations also in Poiseuille flow. 
Accordingly, a perfect agreement between the results
of the numerical simulations and the analytical calculations
on $\phi_p$ and $V_p$  could not be expected. Nevertheless, we
find good agreement between both approaches, cf. Fig.~\ref{F:phi_ratio_pois},
 especially
further away from the center
of the Poiseuille flow, where the linear contribution
in the flow profile dominates and where the assumptions are better fulfilled.

{\it {Acknowledgments}-}
We would like to thank A.~Ziehl and C.~Wagner for inspiring discussions.
This work was supported by the German science foundation
via the priority program on micro- and nanofluidics
SPP 1164 and via the research unit FOR 608.

\begin{appendix}
\section{Cross-correlation $\langle \dev{v}_x(t) \dev{v}_y(t) \rangle$}
\label{App_corr}
In this appendix we discuss the correlation $\langle \dev{v}_x(t) \dev{v}_y(t) \rangle$
and we
show that it includes, besides the contribution $  \propto f_{xy}$ in Eq.~(\ref{E:vxvy}),
 in the limit $k \to 0$ an additional
contribution $\propto f_{yy}$, which corresponds to the 
result described in Ref.~\cite{Drossinos:05}. From Eqs.~(\ref{E:sol_r}) one obtains
\begin{eqnarray}
 \dev{v}_x(t)  &=&  \int_0^t d\tau  a_{vv}(t-\tau)\left[ \frac{F_x^b(\tau)}{m} +
 2 \beta b  ~\dev{y}(\tau)\right] \,,\\
\dev{v}_y(t)  &=&  \int_0^t d\tau a_{vv}(t-\tau) ~\frac{F_y^b(\tau)}{m} \,,\\
\dev{y}(\tau) &=&  \int_0^t d\tau'~ a_{rv}(\tau - \tau') ~ \frac{F_y^b(\tau)}{m}\,, 
\end{eqnarray}
with the abbreviations
\begin{eqnarray}
 a_{vv}(t) &=&  \ez^T ~ e^{\mat{L}(t)}~ \ez \,, \\
 a_{rv}(t)  &=&  \er^T~ e^{\mat{L}(t)}~ \ez \quad \mbox{with} \quad  \er =
\begin{pmatrix}
1 \\ 0
 \end{pmatrix}\,.
\end{eqnarray}
The equal-time velocity correlation takes then the form
\begin{align}
  &\langle \dev{v}_x(t) \dev{v}_y(t) \rangle =\nonumber \\
  & \int_0^t d\tau \int_0^t   d\tau'
a_{vv} (t-\tau) a_{vv}(t-\tau') \langle \frac{F_x^b(\tau) F_y^b(\tau')}{m^2} \rangle
\nonumber \\
 &+ 2 \beta b \int_0^t d\tau a_{vv}(t-\tau) \langle \dev{y}(\tau) \dev{v}_y(t)\rangle\,.
\end{align}
With the correlation between the position $\dev{y}(t)$ and the
velocity $\dev{v}_y(t)$,
\begin{align}
 \langle \dev{y}(\tau) \dev{v}_y(t) \rangle = \frac{f_{yy}}{m^2}
\int_0^{min(t,\tau)} d\tau'' a_{rv}(\tau-\tau'') a_{vv}(t-\tau'')\,,
\end{align}
one ends up  with the expression
\begin{align}
& \langle \dev{v}_x(t) \dev{v}_y(t) \rangle = \frac{f_{xy}}{m^2} 
\int_0^t d\tau a_{vv}(t-\tau)^2 \nonumber \\
 & +2\beta b \frac{f_{yy}}{m^2} \int_0^t d\tau \int_0^\tau 
 d\tilde \tau  ~a_{vv}(t-\tau) a_{rv}(\tau-\tilde \tau) a_{vv}(t-\tilde\tau)
\nonumber \\
  &=  \frac{f_{xy}}{m^2} \frac{G_1(t)}{2}  + 2\beta b \frac{f_{yy}}{m^2}
\frac{G_2(t)}{4}\,,
\label{vxvyG}
\end{align}
where we have introduced the functions $G_1(t)$ and $G_2(t)$ as well as
 the parameter
$\delta$:
\begin{align}
 G_1(t)  &= \left[ \frac{\beta\cos^2(\delta t)}{\delta^2} +\frac{\cos(\delta t)
\sin(\delta t)}{\delta}  \right. \nonumber \\
 &-\left. \frac{\delta^2+2\beta^2}{2\delta^2\beta} \right] ~e^{-2\beta t} +
\frac{1}{2\beta}\,,\\
\label{G2exp}
 G_2(t)  &= \left[ \frac{\sin(\delta t)\cos(\delta t)}{2\delta^3} -\frac{\beta
\sin^2(\delta t)}{\delta^4} +\frac{ t\sin^2(\delta t)}{2\delta^2} \right. \nonumber
\\
 &+\left. \frac{\beta t \cos(\delta t)\sin(\delta t)}{\delta^3} -\frac{t\cos^2(\delta
t)}{2\delta^2}\right] ~e^{-2\beta t}\,,\\
\delta &= \sqrt{\omega^2 -\beta^2}\,.
\end{align}

In the limit $t \to \infty$ the contribution $G_1(t)$ remains finite,
$\lim_{t \to \infty}G_1(t)=\frac{1}{2\beta}$.
For finite values of $\tau_p=\zeta/k$ the function $G_2(t)$ vanishes 
on a timescale $t \gg 1/(2\beta)$ too.
 But in the limit of a vanishing laser tweezer
potential, $k \sim \omega^2 \to 0$, one has $\delta \simeq \pm i(\beta-\omega^2/(2\beta)+...)$ and in this limit some contributions of $G_2(t)$ cancel each other and the remaining terms are proportional to $\exp(-2t/\tau_p)$. These contributions do not decay in the limit $\tau_p \to \infty$ and one gets $G_2(t) = 1/(8\beta^3)$. In this case one obtains even for $f_{xy}=0$ a non-zero cross-correlation 
$\langle \dev{v}_x(t)\dev{v}_y(t) \rangle = \frac{b f_{yy}}{4\zeta^2}$, which is exactly the same result as described in  \cite{Drossinos:05}.

\section{Anisotropic trap}
In experiments the force exerted by the optical tweezer on a colloidal particle may be anisotropic in the shear plane.
For instance, the force constant may be different along and perpendicular to the laser beam.
If the laser beam does not hit the shear plane perpendicularly the difference between the force constants in $x$- and $y$-direction increases with the deviation from the orthogonal direction to the shear plane.
Let's assume in Eq. \eqref{E:springforce} a different force constant for each direction: $k_x$, $k_y$, $k_z$.

The relaxation of the particle is now different along different directions around the potential minimum. 
In order to take this effect into account, we introduce three relaxation times in the same way as in Sec.~\ref{S:fluct},
\begin{align}
	\tau_{p;i} = \frac{2\beta}{\omega_i^2} = \frac{\zeta}{k_i}\,,
\end{align}
and we further define the fraction
\begin{align}
\mu := \frac{\omega_y}{\omega_x} = \frac{k_y}{k_x}.
\end{align}
between the eigenfrequencies in the shear plane, which is used to express $k_y$ by $k_x$ via $k_y = \mu k_x$.

The covariance matrix $\mat{C}$ from Sec.~\ref{S:fluct} depends now in a complex manner on
the different force constants, but in the limit $\mu \to 1$ and $k_z=k_x$ the result from the previous sections are obtained again.

The different static correlations as given for the isotropic forces in Sec. \ref{S:fluct} 
change as follows:
\begin{align}
\langle \dev{v}_y^2 \rangle &= \frac{1}{4}\frac{f_{yy}}{m^2 \beta}\,,
\label{E:vy2b}
\\
\langle \dev{y}^2 \rangle &= \frac{1}{4}\frac{f_{yy}}{m^2 \beta \omega_y^2} 
		     = \frac{\langle \dev{v}_y^2 \rangle}{\omega_y^2}\,,
\label{E:ry2b} \\
\langle \dev{v}_x^2 \rangle
        &=  \frac{1}{4}\frac{f_{xx}}{m^2\beta}
		+ \frac{1}{2} b^2 \langle \dev{y}^2 \rangle \Theta_1
		+ \frac{8}{3} c^2 \frac{ \omega_x^2 \tau_{p;x}^2 }
				  {(1 + 2 \tau_{p;x}^2 \omega_x^2)} 
			\langle \dev{y}^2\rangle^2 \Theta_2 \notag\\
 	&\quad -b \frac{1-\mu^2}{1+\mu^2} 
	\frac{2 \langle \dev{v}_x \dev{v}_y \rangle}{\tau_{p;x}\omega_x^2}
	\,, \label{E:vx2b} \\
\langle \dev{x}^2 \rangle
        &=   \frac{\langle \dev{v}_x^2 \rangle}{\omega_x^2} 
	   + \frac{1}{2} \tau_{p;x}^2 b^2 \langle \dev{y}^2 \rangle \Pi_1
	   + \frac{2}{3} \tau_{p;x}^2 c^2 \langle \dev{y}^2 \rangle^2 \Pi_2
\notag \\
	   &\qquad\qquad\qquad + b \tau_{p;x} 
	\frac{2 \langle \dev{v}_x \dev{v}_y \rangle}{\omega_x^2(1+\mu^2)}
	   \,, \label{E:rx2b}  \\
\langle \dev{v}_z^2 \rangle &= \frac{1}{4}\frac{f_{zz}}{m^2 \beta}\,,
\label{E:vz2b} \\
\langle \dev{z}^2 \rangle &= \frac{1}{4}\frac{f_{zz}}{m^2 \beta \omega_y^2} 
		     = \frac{\langle \dev{v}_z^2
\rangle}{\omega_z^2}\,, \label{E:rz2b}
\end{align}
where the following abbreviations have been introduced:
\begin{align}
	 \Theta_1 &= 2 \frac{\mu^2}{1+\mu^2}\,, \notag \\
	 \Theta_2 &= 3 \mu^2 \frac{\tau_{p;x}^2 \omega_x^2 
				 +\frac{1}{6} (1+2\mu^2)}
	 		        {(1+2\mu^2)\tau_{p;x}^2 \omega_x^2  
				+ \frac{1}{6}(1-4\mu^2)^2} \,,\notag \\
	\Pi_1 &= \displaystyle{ 2 \frac{(1+\mu^2)^2 + \frac{(1-\mu^2)(1+2\mu^2)}{{\tau_{p;x}}^2 \omega_x^2}
	                 + \frac{1}{4}\frac{(1-\mu^2)^3}{{\tau_{p;x}}^4 \omega_x^4}
		      }
		      {  (1+\mu^2)\Big((1+\mu^2)+\frac{1}{2}
\frac{1}{{\tau_{p;x}}^2 \omega_x^2}(1-\mu^2)^2\Big)^2
		      } }\,,\notag \\
	\Pi_2 &= \frac{1 + \frac{1}{3} \frac{1}{{\tau_{p;x}}^2 \omega_x^2}
(2+\mu^2) 
			 + \frac{1}{12} \frac{1}{{\tau_{p;x}}^4 \omega_x^4}
(1-2\mu^2)(1-4\mu^2)
		      }
		      {1 + \frac{1}{2} \frac{1}{{\tau_{p;x}}^2 \omega_x^2}
		      	   \left(
		      		\frac{(1+2\mu^2)}{3} 
					+ \frac{1}{2}
\frac{1}{{\tau_{p;x}}^2 \omega_x^2}
					  \frac{(1-4\mu^2)^2}{9}
		      	   \right)}\,. \notag
\end{align}

Since the $y$- and $z$-displacements are independent of any other direction, as can be
seen in Eqs. \eqref{E:sol_r}, their autocorrelations are functions of the individual force
constant only and do not depend on $\mu$. However, the autocorrelations for the velocity and the position in $x$-direction
depends in a complex manner on the different force constants.
The same applies for the cross-correlations in the shear plane  as follows
\begin{align}
\langle \dev{v}_x \dev{v}_y \rangle &= 
\langle \dev{v}_y \dev{v}_x \rangle \label{E:vxvy2b} \\
	&= \frac{1}{4} \frac{f_{xy}}{m^2\beta} 
		\frac{\tau_{p;x}^2 \omega_x^2 (1+\mu^2)}
     	{\bigg(\tau_{p;x}^2 \omega_x^2 (1+\mu^2) 
	+ \frac{1}{2}(1-\mu^2)^2 \bigg)} \notag \\
&\quad + \frac{1}{4}b \langle \dev{y}^2 \rangle 
	\frac{\tau_{p;x} \omega_x^2 \mu^2(1-\mu^2)}
	{\bigg(\tau_{p;x}^2\omega_x^2(1+\mu^2) 
	+ \frac{1}{2}(1-\mu^2)^2 \bigg)}
 \notag \\
\langle \dev{x} \dev{y} \rangle &= \langle \dev{y} \dev{x} \rangle
	= \bigg[\frac{\langle \dev{v}_x\dev{v}_y \rangle}{\omega_x^2} 
	   + \frac{1}{2} \tau_{p;x} b \langle \dev{y}^2 \rangle \bigg]
		\frac{2}{(1+\mu^2)} \label{E:xy2b} \\ 
\langle \dev{x} \dev{v}_y \rangle &= -\langle \dev{y} \dev{v}_x \rangle
	= -\frac{\mu^2 b \langle \dev{y}^2 \rangle}{(1+\mu^2)}
	  + \frac{(1-\mu^2)}{(1+\mu^2)} \frac{\langle \dev{v}_x\dev{v}_y
\rangle}{\tau_{p;x}\omega_x^2} \,.
	    \label{E:xvy2b}
\end{align}

\end{appendix}


\begin{thebibliography}{10}


\bibitem{Einstein:1905.1}
A. Einstein, Ann. Phys. {\bf 17},  549  (1905).

\bibitem{Dhont:96}
J.~K.~G. Dhont, {\em {An Introduction to Dynamics of Colloids}} (Elsevier,
  Amsterdam, 1996).

\bibitem{Stone:2001.1}
H. Stone and S. Kim, {AIChe J.} {\bf {47}},  {1250}  ({2001}).

\bibitem{Ottino:2004.1}
J. Ottino and S. Wiggins, {Phil. Trans. R. Soc. Lond. A} {\bf {362}},  {923}
  ({2004}).

\bibitem{Holzer:2006.1}
L. Holzer and W. Zimmermann, Phys. Rev. E {\bf {73}},  {060801(R)}  ({2006}).

\bibitem{Bechinger:2006.1}
C. Lutz, M. Reichert, H. Stark, and C. Bechinger, Europhys. Lett. {\bf 74},
  719  (2006).

\bibitem{TaylorGI:1953.1}
G. Taylor, {Proc. R. Soc. London A} {\bf {219}},  {186}  ({1953}).

\bibitem{Steinberg:2001.2}
A. Groisman and V. Steinberg, {Nature} {\bf {410}},  {905}  ({2001}).

\bibitem{Chu:1997.1}
T. Perkins, D. Smith, and S. Chu, {Science} {\bf {276}},  {2016}  ({1997}).

\bibitem{deGennes:1997.1}
{P.~G.~deGennes}, {Science} {\bf {276}},  {1999}  ({1997}).

\bibitem{Steinberg:2000.1}
A. Groisman and V. Steinberg, {Nature} {\bf {405}},  {53}  ({2000}).

\bibitem{SGrossmann:2000.1}
S. Grossmann, {Rev. Mod. Phys} {\bf {72}},  {603}  ({2000}).

\bibitem{Eckhardt:2003.1}
B. Eckhardt and R. Pandit, Eur. Phys. J. B {\bf {33}},  {373}  ({2003}).

\bibitem{Oberlack:2006.1}
G. Khujadze, M. Oberlack, and G. Chagelishvili, Phys. Rev. Lett. {\bf {97}},
  034501  ({2006}).

\bibitem{Ziehl:2009.1}
A. Ziehl {\it et~al.}, {Phys. Rev. Lett.}  {\bf {103}}, {230602}  ({2009})

\bibitem{Bedeaux:1995.1}
K. Miyazaki and D. Bedeaux, {Physica A} {\bf {217}},  {53}  ({1995}).

\bibitem{Brady:04}
G. Subramanian and J. Brady, {Physica A} {\bf {334}},  {343}  ({2004}).

\bibitem{Drossinos:05}
Y. Drossinos and M.~W. Reeks, Phys. Rev. E {\bf {71}},  {031113}  ({2005}).

\bibitem{Sancho:1979.1}
M. San~Miguel and J.~M. Sancho, Physica {\bf {99A}},  {357}  ({1979}).

\bibitem{Holek:01}
I. Santamaria-Holek, D. Reguera, and J.~M. Rubi, Phys. Rev. E {\bf 63},
  {051106}  (2001).

\bibitem{vdVen:1980.1}
R. Foister and T. {van de Ven}, J. Fluid Mech. {\bf {96}},  {105}  ({1980}).

\bibitem{Leal:1980.1}
G. Fuller, J. Rallison, R. Schmidt, and L. Leal, J. Fluid Mech. {\bf {100}},
  {555}  ({1980}).

\bibitem{Holzer:2009.2}
L. Holzer, Ph.D. thesis, {Universit\"at Bayreuth}, {2009}.

\bibitem{vdWater:1998.1}
M. Hoppenbrouwers and W. {van de Water}, Phys. Fluids {\bf {10}},  {2128}
  ({1998}).

\bibitem{Grier:2003.1}
D.~G. Grier, Nature {\bf 424},  810  (2003).

\bibitem{Ashkin:1986.1}
A. Ashkin, J. Dziedzic, J. Bjorkholm, and S. Chu, Opt. Lett. {\bf {11}},  {288}
   ({1986}).

\bibitem{Chu:1991.1}
S. Chu, {Science} {\bf {253}},  {861}  ({1991}).

\bibitem{Chu:1994.2}
T. Perkins, S. Quake, D. Smith, and S. Chu, {Science} {\bf {264}},  {822}
  ({1994}).

\bibitem{Chu:1995.1}
T. Perkins, D. Smith, R. Larson, and S. Chu, {Science} {\bf {268}},  {83}
  ({1995}).

\bibitem{Brochard:1995.1}
F. Brochard-Wyart, Europhys. Lett. {\bf {30}},  {387}  ({1995}).

\bibitem{Larson:1997.1}
R.~G. Larson, T.~T. Perkins, D.~E. Smith, and S. Chu, Phys. Rev. E {\bf {55}},  {1794}
  ({1997}).

\bibitem{Rzehak:99.2}
R. Rzehak, D. Kienle, T. Kawakatsu, and W. Zimmermann, Europhys. Lett. {\bf
  {46}},  {821}  ({1999}).

\bibitem{Rzehak:00.1}
R. Rzehak, W. Kromen, T. Kawakatsu, and W. Zimmermann, Eur. Phys. J. E {\bf
  {2}},  {3}  ({2000}).

\bibitem{Kienle:01.1}
D. Kienle and W. Zimmermann, {Macromolecules} {\bf {34}},  {9173}  ({2001}).

\bibitem{Rzehak:02.1}
R. Rzehak and W. Zimmermann, Europhys. Lett. {\bf {59}},  {779}  ({2002}).

\bibitem{Quake:99.1}
J.~C. Meiners and S.~R. Quake, Phys. Rev. Lett. {\bf {82}},  {2211}  ({1999}).

\bibitem{Bartlett:2002.1}
S. Henderson, S. Mitchell, and P. Bartlett, Phys. Rev. Lett. {\bf {88}},
  088302  ({2002}).

\bibitem{Grier:2001.1}
E. Dufresne, D. Altman, and D. Grier, Europhys. Lett. {\bf {53}},  {264}
  ({2001}).

\bibitem{Grier:2000.1}
E.~R. Dufresne, T.~M. Squires, M.~P. Brenner, and D.~G. Grier, Phys. Rev. Lett. {\bf {85}},
   {3317}  ({2000}).

\bibitem{Schmidt:2005.1}
M. Atakhorrami, G.~H. Koenderink, C.~F. Schmidt, and F.~C. MacKintosh, Phys. Rev. Lett.
  {\bf {95}},  208302  ({2005}).

\bibitem{Weitz:2000.1}
J. Crocker {\it et~al.}, Phys. Rev. Lett. {\bf {85}},  {888}  ({2000}).

\bibitem{Quake:2006.1}
M. Polin, D.~G. Grier, and S. Quake, Phys. Rev. Lett. {\bf {96}},  088101
  ({2006}).

\bibitem{Grier:2002.1}
P.~T. Korda, M.~B. Taylor, and D.~G. Grier, Phys. Rev. Lett. {\bf 89},  128301
  (2002).

\bibitem{Bammert:2008.1}
J. Bammert, S. Schreiber, and W. Zimmermann, Phys. Rev. E {\bf 77},  042102
  (2008).

\bibitem{Bammert:2009.1}
J. Bammert and W. Zimmermann, Eur. Phys. J. E {\bf 28},  331  (2009).

\bibitem{Dholakia:2003.1}
M.~P. MacDonald, G.~C. Spalding, and K. Dholakia, Nature {\bf 426},  421
  (2003).

\bibitem{Simons:BPJ70-96-1813}
R. Simmons, J. Finer, S. Chu, and J. Spudich, Biophys. J. {\bf {70}},  {1813}
  ({1996}).

\bibitem{Crocker:JCIS179-96-298}
J. Crocker and D. Grier, J. Colloid Int. Sci. {\bf {179}},  {298}  ({1996}).

\bibitem{LanLifVI}
L.~D. Landau and E.~M. Lifschitz, {\em {Lehrbuch der Theoretischen Physik:
  Hydrodynamik}}, 2nd  ed. (Akademie Verlag, Berlin, 1987).

\bibitem{Stokes:1850.1}
G.~G. Stokes, {Trans. Cambridge Phil. Soc.} {\bf IX},  8  (1850).

\bibitem{Faxen:27.1}
H. Faxen, {Z. angew. Math. Mech.} {\bf 7},79    (1927).

\bibitem{Reichl:1998}
L.~E. Reichl, {\em A Modern Course in Statistical Physics}, 2. ed. (Wiley-VCH,
  Berlin, 1998).

\bibitem{LanLifIX}
L.~D. Landau and E.~M. Lifschitz, {\em {Lehrbuch der Theoretischen Physik:
  Statistische Physik II}} (Akademie Verlag, Berlin, 1980).

\bibitem{Kampen}
{N.~G. van~Kampen}, {\em {Stochastic Processes in Physics and Chemistry}}
  (Elsevier, Amsterdam, 2004).

\bibitem{Bamieh:2001}
B. Bamieh and M. Dahleh, Phys. Fluids {\bf 13},  3258  (2001).

\bibitem{zarate:2008}
J.~M. {Ortiz de~Z\'arate} and J.~V. Sengers, Phys. Rev. E {\bf 77},  026306
  (2008).

\bibitem{Kraus:96}
M. Kraus, W. Wintz, U. Seifert, and R. Lipowsky, Phys. Rev. Lett. {\bf {77}},
  {3685}  ({1996}).

\bibitem{deGennes:2006}
{P.~G.~deGennes, J.~Prost}, {\em {The Physics of Liquid Crystals}} (Clarendon
  Press, Oxford, 2006).

\bibitem{Misbah:2009.1}
G. Danker, P.~M. Vlahovska, and C. Misbah, Phys. Rev. Lett. {\bf {102}},
  {148102}  ({2009}).



\end{thebibliography}
\end{document}